# Towards Plug-and-Play Protection for Meshed Distribution Systems with DG

Aristotelis M. Tsimtsios, *Student Member, IEEE*, and Vassilis C. Nikolaidis, *Senior Member, IEEE*

*Abstract*--Future distribution systems are expected to display increased complexity, mainly due to looped/meshed operation, switch between grid-connected and islanded mode and considerable integration of distributed generation. This paper investigates a plug-and-play protection solution for overhead distribution systems with such variable operation conditions, employing existing numerical relay capabilities. This solution is applied by designing plug-and-play, communication-assisted, multifunctional relays with integrated protection element settings, which apply universally to all distribution system conditions, rendering the protection scheme independent of a specific system. Hence, the need for user-defined settings or future revisions due to system changes is eliminated. The scheme ensures coordination between main line relays and backup protection of laterals, without a coordination study. There is no need to replace or modify existing lateral protection means for this purpose; only their known time-overcurrent curves are uploaded to the relays by the user. The scheme is described and evaluated through simulations in two test systems. Meaningful conclusions are finally derived.

*Index Terms*--Distributed generation, distribution system, meshed network, plug-and-play protection.

## I. INTRODUCTION

FUTURE distribution systems are envisioned including a considerable amount of distributed generation (DG), operating in a looped or meshed network configuration (i.e. in a looped configuration, also including interconnection line segments) and freely switching between grid-connected (GC) and islanded (ISL) mode. Among others, advanced protection schemes are needed for such a complex system operation.

Determining proper relay settings is a challenging task. In fact, besides technical reasons, a common cause of protection maloperation is incorrect relay setting [1]. Even in conventional overhead (OH) radial distribution systems, setting the main line relay(s) requires a cumbersome simulation study, mainly due to the need for coordination with lateral protection means. Obviously, the difficulty increases considerably in looped/meshed distribution networks with DG, due to the increased network's complexity and the need for frequent relay setting revisions, given the continuous expected changes (e.g. connection of new DG units). Hence, it becomes a necessity to investigate reliable protection concepts, which eliminate the need for simulation-based user-defined settings, being, as far as possible, independent of a specific network. Distribution system particularities (e.g. existence of laterals) and existing relay capabilities should also be considered. To the best of the authors' knowledge, so far, papers focusing on protection of looped/meshed distribution systems have not addressed a "leave-and-forget" concept including all the above attributes.

Several papers focusing on the protection of looped or meshed distribution systems with DG apply directional overcurrent relays (DOCRs). Optimization algorithms have been recently proposed for setting DOCRs [2]-[4]. In [2], DOCRs are optimally set for both the GC and the ISL mode of system operation, without using communication means. In [3], [4], communication-assisted protection schemes are applied, using dual-setting DOCRs. The latter are optimally set, while, blocking signals are used to ensure selectivity. However, to produce optimal settings, these algorithms require data extracted from the specific system and/or system conditions considered. Also, they do not address coordination with lateral protection.

Communication/pilot-based DOCR schemes have been also proposed in [5]-[8]. The authors of [5] apply DOCRs employing blocking and inter-trip signals, to ensure selectivity and isolation of the faulted line segment, respectively. The extension of this scheme to looped and meshed networks is also discussed. DOCRs are applied to protect a looped distribution network with DG in [6], based on a permissive logic. In [7], [8], permissive/blocking-based DOCR schemes are considered for real looped distribution networks. Proper relay setting is required in [5]-[8], based on the specific system to protect.

Differential protection has been also considered for looped/meshed distribution systems with DG [9]-[11]. To ensure both sensitivity and security, proper setting of differential relays is required, especially when loads and/or DG units are connected within the differential protection zone [11], [12]. Especially in [11], where the proposed multi-agent differential scheme must coordinate with the lateral fuses, the relays must be further set for this purpose. Also, a drawback of commercial differential relays is their inability to inherently protect network-parts outside of their primary protection zone [12].

In [13], a communication-assisted multi-agent protection scheme is proposed, using current phase angle comparison. To define the angle setting, an extensive simulation study is needed. Coordination with lateral protection is also not addressed.

Communication/pilot-based distance protection schemes for looped/meshed distribution systems with DG are proposed in [14], [15]. The scheme of [14] mainly applies a permissive logic, while, it also uses blocking signals only against weak-

The research work was supported by the Hellenic Foundation for Research and Innovation (HFRI) and the General Secretariat for Research and Technology (GSRT), under the HFRI PhD Fellowship grant (GA. no. 19773).
A. M. Tsimtsios and V. C. Nikolaidis are with the Department of Electrical and Computer Engineering, Democritus University of Thrace, Xanthi 67100, Greece (e-mails: atsimtsi@ee.duth.gr; vnikolai@ee.duth.gr).

infeed conditions. An extensive simulation study in the protected network is required to set the distance relays. In [15], both a pilot-based (using permissive and blocking signals) and a central distance protection scheme are proposed. Proper setting is required for the distance relays, while, coordination with common lateral protection means is not addressed.

Multifunctional protection schemes for distribution systems with DG, using communication means, are proposed in [16] (employing, among others, directional-overcurrent, undervoltage and high-impedance-fault-detection functions) and [17] (employing differential and directional-overcurrent functions). Proper relay settings have to be determined for these schemes, especially for backup protection coordination purposes.

Communication-assisted adaptive protection schemes have been also proposed for looped/meshed distribution systems with DG, employing different setting groups [18]-[20]. In [18], [19], each setting group of the applied DOCRs is defined in advance using optimization algorithms. Adaptive distance relays are considered in [20]. The main challenge in these approaches is that they require offline load-flow/short-circuit studies in the protected network, considering all the possible configurations. A solution to optimally determine the required number of setting groups for each DOCR is proposed in [21]. A different adaptive approach is followed in [22], which calculates the new DOCR settings based on an initial setting group, stored in the relays in advance. The initial settings are calculated based on the specific system to protect. Multi-agent adaptive approaches are proposed in [23], [24], while, online adaptive protection is applied in [25]. These schemes have to be set/designed based on data extracted from the specific system protected. Moreover, none of [18]-[25] addresses coordination with existing lateral protection means.

Finally, several research efforts examine alternative protection solutions for looped/meshed distribution networks, not based on commercial relay functions. Protection techniques applying data-mining [26], [27], machine learning [28] and deep neural networks [29] have been proposed among others, which, however, require a training process, based on scenarios of the specific network examined. Other techniques regard dynamic-state-estimation-based protection [30], interval type-2 fuzzy logic [31] and checking power direction of the positive-sequence fault component, as part of a blocking-signal-based pilot protection scheme [32]. In [26]-[32], coordination with existing lateral protection means is not addressed.

To sum up, the protection schemes proposed so far for looped/meshed distribution systems are characterized by one or more of the following drawbacks:
i) They require extensive relay setting studies, taking into account any possible fault and system configuration scenario. As mentioned before, this might be complex in looped/meshed distribution systems with DG.
ii) They are set/designed based on data of the specific network to protect, which renders them vulnerable to unplanned system changes (e.g. new DG connection, change in the short-circuit capacity of the external grid, etc.).
iii) Either they do not address coordination between the main line relays and existing lateral protection means, or they

require properly setting the relays to achieve coordination.

To address all the above drawbacks, eliminating protection system design complexity in modern OH distribution systems, a plug-and-play (PnP) protection solution is investigated in this work. The main contribution/novelty of this solution, compared to other proposed protection solutions, is reflected on the concurrent fulfillment of the following:

- The PnP relays do not need user-defined settings, typically resulting from a simulation study. Instead, they are designed by default to deal with any fault/system scenario.
- The protection scheme is independent of a specific OH distribution system and immune to planned or unplanned system changes, such as switch between GC and ISL mode, connection/disconnection of DG units or line segments etc. Hence, applying adaptive logic is not needed, while, the need for future setting revisions is eliminated.
- The PnP relays operate selectively with each other and with the existing lateral protection (even non-settable fuses), without needing a coordination study. Especially the latter is automatically achieved, by simply uploading the characteristics of lateral protection means to the relays.

It should be mentioned that the proposed scheme addresses non-resistive/resistive faults of any type/location. Also, the PnP relays are intentionally designed by properly combining existing protection techniques, applicable with commercial relay technology. This is to enhance the schemes' practicality. The novelty of the proposed solution is also reflected on the way that these individual techniques are combined.

This paper is organized as follows. The basic logic of the proposed scheme and its design using PnP relays is described in Section II and III, respectively, while, its performance is evaluated in Section IV. Section V includes a comparison with differential protection. Conclusions are drawn in Section VI.

## II. BASIC LOGIC OF THE PROTECTION SCHEME

The PnP protection scheme is basically described via Fig. 1, showing a part of a generic OH meshed distribution system. Note that the scheme's logic, addressed below, ensures isolation of the faulted line part, even if the network-loop is open.

### A. Main Line Segment Protection

Each main line segment $L_{i,j}$, connecting buses $B_i$ and $B_j$, is protected by two PnP, communication-assisted, multifunctional, numerical relays $R_{i,j}$ and $R_{j,i}$, installed at its opponent ends, which supervise for forward faults (i.e. fault currents flowing into $L_{i,j}$). A protection security measure is necessary here, to ensure that none of these relays will undesirably trip for a main line fault outside of $L_{i,j}$. A reliable option would be to apply a directional-comparison-blocking (DCB)-based logic [1]. According to this logic, each of $R_{i,j}$ and $R_{j,i}$, would block its opponent relay in case of detecting a reverse fault (i.e. outside of $L_{i,j}$). An advantage of such a logic (e.g. compared to the permissive one) is its operability even if the looped configuration is lost [7]. Although such schemes are typically applied to transmission lines, the use of blocking signals for selectivity purposes has been also proposed for distribution systems [3]-[5], [7], [8], [14], [15], [32]. In these applications, blocking signals are transmitted by the relays, once they detect a fault in a specific direction. The particularity of this work compared to other relevant applications, as well as the traditional DCB logic, is that a relay does not decide on whether to block another relay, after determining fault direction. Instead, each of $R_{i,j}$ and $R_{j,i}$ sends a blocking signal (signals *bf* in Fig. 1) to the forward element of its opponent relay, immediately after detecting a probable fault situation. Only if, after at least one fundamental period of processing (i.e. starting, phase selection and fault direction determination), this situation is found to be a forward fault (i.e. a fault inside $L_{i,j}$), *bf*-signal is cancelled, and the opponent relay is allowed to instantaneously trip its circuit breaker ($CB_{i,j}$ and $CB_{j,i}$, respectively). In this way, it is ensured that a relay will be timely blocked in case of an external fault, without the need to consider an intentional time delay in its operation, to allow for blocking signal transmission (as it is a common practice [7], [32]). Finally, once a relay trips its *CB*, it inter-trips the opponent *CB* as well. Note that the inter-trip order is redundant and it is issued to ensure fault clearance, in the case where one of the two opponent relays fails to send a trip order to its *CB* or it cannot sense the fault. The latter regards either weak-infeed conditions or loss of the looped configuration.

*B. Bus Protection*

For bus protection, we exploit the already installed adjacent relays. For instance, bus $B_i$ in Fig. 1 is protected by $R_{i,j}$, $R_{i,h}$ and $R_{i,l}$, which, besides supervising for forward faults to protect $L_{i,j}$, $L_{h,i}$ and $L_{i,l}$, respectively, they also supervise for reverse faults to protect $B_i$. These relays are assumed forming a group of relays $GR_i$. To enhance protection security as previously, each $GR_i$-relay sends a blocking signal (signals *br* in Fig. 1) to the reverse element of the rest $GR_i$-relays, immediately after detecting a probable fault situation. If, ultimately, this situation is found to be a reverse fault, *br*-signal is cancelled. Hence, the rest $GR_i$-relays are let trip their assigned *CB* instantaneously. In this case too, once a $GR_i$-relay trips its *CB*, it also inter-trips the *CBs* of the rest $GR_i$-relays ($CB_i$).

*C. Lateral Protection*

Each lateral $L_i$ is primarily protected by the protection means $P_i$ installed at its departure. $P_i$ is usually a fuse, due to its relatively low cost, and, less often, a conventional overcurrent relay. In this work, existing $P_i$ is maintained, so as to avoid $P_i$-replacement cost. We also assume that $P_i$ always co-ordinates with its downstream protection means. Besides providing primary protection to the main line segments and buses, the proposed scheme also aims to provide backup protection for each lateral $L_i$. This task is assigned to the $GR_i$-relays, using the same basic logic as that used for $B_i$ protection. However, $GR_i$-relays must further coordinate with $P_i$. Research efforts which address the issue of coordination between the main line relays and common lateral protection means in looped/meshed distribution systems (e.g. [11], [14]) require setting the main line relays properly for this purpose. The latter is not required by the proposed scheme, thanks to the internal protection logic described in the next section.

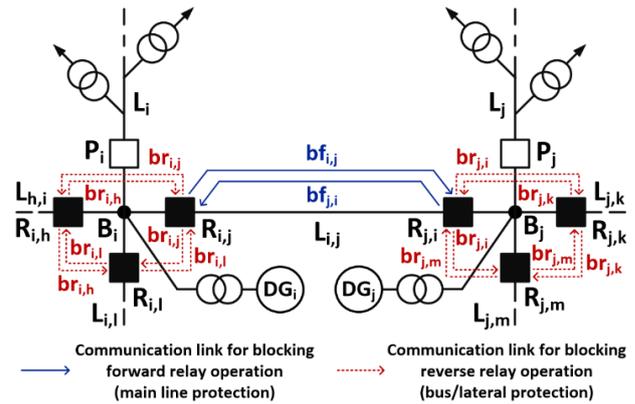

Fig. 1. Basic logic of the proposed protection scheme.

### III. DESIGN OF THE PROPOSED PLUG-AND-PLAY RELAY

In the following, the functional logic of a designed PnP relay $R_{i,j}$ is described, and is illustrated in Fig. 2. A dedicated element is designed against three-phase (3PH), double-phase (2PH), and ground (double-phase-ground (2PHG) and single-line-ground (SLG)) faults. Two additional elements are designed for addressing reverse bus/lateral faults and backup protection of main line segments and buses. The basic-calculations block processes the relay measurements and generates signals which are used as an input to the rest elements. Fig. 2 also indicates the subsections of the paper where each element or element-part is discussed.

*A. Ground (SLG/2PHG)-Fault Protection Element*

*1) Starting*

To initiate the ground-fault protection element, preserving adequate sensitivity, an overcurrent starting function based on superimposed quantities is used. In general, the superimposed quantity $n_s$, of a quantity $n$ at instant $t$, is calculated as [33]:

$$n_s(t) = n(t) - n(t - kT) \quad (1)$$

where $k$ is an integer (here $k = 3$) and $T$ is the fundamental period. $n_s$ reflects a change in $n$ and equals (almost) zero under pre-fault conditions, while, it gives the pure fault signal during a fault.

The starting function of the ground-fault protection element asserts (see Fig. 2) when the magnitude of the superimposed zero-sequence current ($I_{0,s}$) satisfies the following condition:

$$3 \cdot I_{0,s} \geq 5\% I_n \quad (2)$$

In (2), $3 \cdot I_{0,s}$ is the superimposed residual current; the latter is chosen here for the starting criterion, as it characterizes unbalanced ground faults. $I_n$ is the nominal current of the relay's current transformer (CT) and $5\% I_n$ is a default overcurrent threshold. This threshold corresponds to a typical maximum current sensitivity, adopted by several commercial relays nowadays (e.g. [34]). The default $5\% I_n$ threshold has been chosen so as to sensitize the starting function as much as possible, keeping up with the capabilities of commercial relays. Note that this threshold is not binding; if a lower current sensitivity can be achieved by a relay manufacturer, a lower default threshold can be defined in advance. Despite sensitization, protection security is ensured through the proposed protection and communication logic (refer to the previous section).

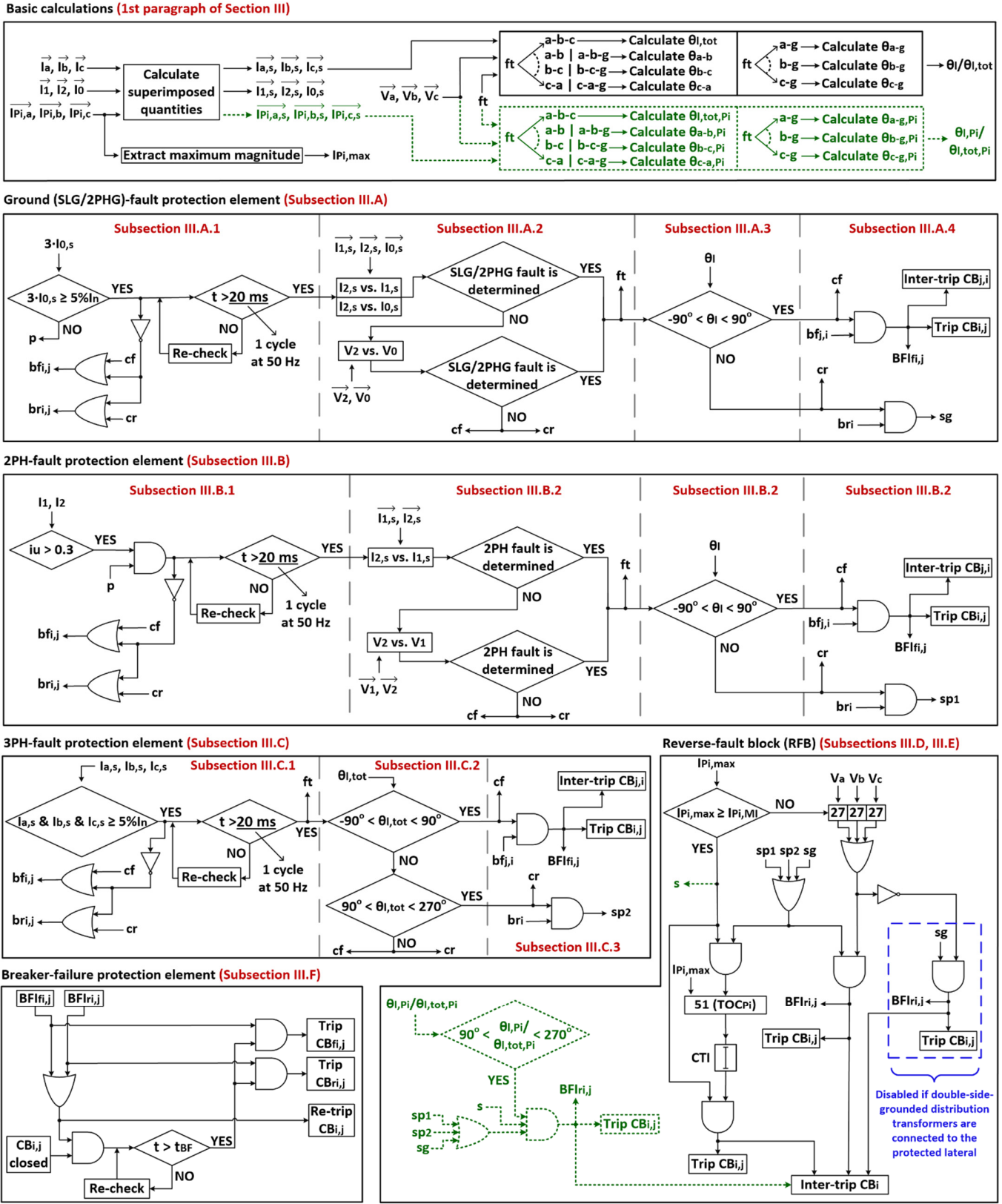

Fig. 2. Overall internal protection and communication logic of a PnP relay $R_{i,j}$ designed for the proposed scheme.

Equation (2) is applicable no matter the mode of system operation (i.e. GC vs. ISL), and even if single-phase DG units are connected to the network, on condition that the main substation transformer or the step-up transformer of the DG units is grounded at the medium-voltage (MV) side. This is because, $3 \cdot I_{0,s}$ will always appear in case of unbalanced ground faults. Dealing with erroneous ground fault detection due to unbalanced loads is discussed later in Subsection III.E.

An exception regarding the applicability of (2) might appear in case of isolated/compensated networks. However, in these networks, a very small earth current flows during SLG faults. In fact, the fault may self-extinguish, or the system may operate with the continuous earth current flowing, without disruption to consumers. Hence, protection means should not undesirably trip in this case anyway; instead, special methods should be used just for fault detection [35]. In case of isolated/compensated networks, the ground-fault protection element is not used at all. 2PHG faults can be dealt with by the 2PH-fault protection element (seen as 2PH faults), described later.

Assuming a full-cycle Discrete Fourier Transform (DFT) filter, reliable enough signals are obtained 1 cycle after the initiation. Hence, indicatively assuming a system frequency of 50 Hz, the overall starting (response) time $t_S$ equals:

$$t_S = t_D + \underbrace{20 \text{ ms}}_{\text{1 cycle at 50 Hz}} \quad (3)$$

where $t_D$ is the time needed for the starting condition to become valid, after fault inception. All the signals used in the algorithm-steps 2) and 3), described next, are those measured/calculated at $t = t_S$. Note that once (2) becomes valid, $bf_{i,j}$ and $br_{i,j}$ are sent to block the forward operation of $R_{j,i}$ and the reverse operation of the rest $GR_i$-relays, respectively.

*2) Phase selection (PS)*

Next, the fault type/faulted phase is determined by examining the phase-angle relationship between the superimposed negative- ($\overrightarrow{I_{2,s}}$) and positive-sequence ($\overrightarrow{I_{1,s}}$) current phasors (hereafter referred to as $I_{2,s}$ vs. $I_{1,s}$ principle), as well as that between the negative- and zero-sequence ($\overrightarrow{I_{0,s}}$) current phasors (hereafter referred to as $I_{2,s}$ vs. $I_{0,s}$ principle). As these principles are based on the phase-angle difference between sequence-current phasors, they are independent of the exact position of each individual phasor (which varies depending on the fault/system conditions). The above principles have been used for many years in commercial relays and are considered reliable under various fault/system conditions [35], [36].

Table I shows the range of phase-angle difference $\varphi_{Is,21}$ between $\overrightarrow{I_{2,s}}$ and $\overrightarrow{I_{1,s}}$, as well as that of phase-angle difference $\varphi_{Is,20}$ between $\overrightarrow{I_{2,s}}$ and $\overrightarrow{I_{0,s}}$, each constituting a signature of a specific fault type [36], [37]. For security, both PS principles must agree on the fault-type for the latter to be specified. In that way, we also achieve fault-type distinction in $I_{2,s}$ vs. $I_{0,s}$ (e.g. 90° < $\varphi_{Is,20}$ < 150° could mean either a-b-g or c-g fault).

TABLE I
SEQUENCE-CURRENT AND SEQUENCE-VOLTAGE PS PATTERNS

| Fault type | $I_{2,s}$ vs. $I_{1,s}$ $\varphi_{Is,21}$ | $I_{2,s}$ vs. $I_{0,s}$ $\varphi_{Is,20}$ | $V_2$ vs. $V_1$ $\varphi_{V,21}$ | $V_2$ vs. $V_0$ $\varphi_{V,20}$ |
|---|---|---|---|---|
| a-b   | 45° – 75°       | -              | 180° – 300°     | - |
| b-c   | 165° – 195°     | -              | (-60°) – 60°    | - |
| c-a   | (-75°) – (-45°) | -              | 60° – 180°      | - |
| a-b-g | 45° – 75°       | 90° – 150°     | 180° – 300°     | 30° – 150° |
| b-c-g | 165° – 195°     | (-30°) – 30°   | (-60°) – 60°    | (-90°) – 30° |
| c-a-g | (-75°) – (-45°) | 210° – 270°    | 60° – 180°      | 150° – 270° |
| a-g   | (-15°) – 15°    | (-30°) – 30°   | 150° – 270°     | (-90°) – 30° |
| b-g   | 105° – 135°     | 210° – 270°    | (-90°) – 30°    | 150° – 270° |
| c-g   | 225° – 255°     | 90° – 150°     | 30° – 150°      | 30° – 150° |

Some commercial relays switch to sequence-voltage-based PS, if sequence-current-based PS fails [36]. In this work, if sequence-current-based PS fails, PS patterns [38] based on the actually measured sequence-voltage phasors are applied (see Fig. 2), which, in fact, are reliable (among others) in microgrids, even including photovoltaic (PV) units. Actually, the voltage-based PS is used as backup for the current-based PS. To this end, Table I also gives the range of phase-angle difference $\varphi_{V,21}$ between the negative- ($\overrightarrow{V_2}$) and positive-sequence ($\overrightarrow{V_1}$) voltage phasors, as well as that of phase-angle difference $\varphi_{V,20}$ between the negative- and zero-sequence ($\overrightarrow{V_0}$) voltage phasors, each indicating a specific fault type. Hereafter, these principles are referred to as $V_2$ vs. $V_1$ and $V_2$ vs. $V_0$, respectively. Note that $V_2$ vs. $V_0$ is used against ground faults, whereas, $V_2$ vs. $V_1$ is used against 2PH faults (described later). To discriminate between a SLG and a 2PHG fault when using $V_2$ vs. $V_0$ (e.g. 30° < $\varphi_{V,20}$ < 150° could mean either a-b-g or c-g fault), the voltage drop in each phase ($\Delta U_a$, $\Delta U_b$ and $\Delta U_c$) is examined. For instance, in the above example, if both $\Delta U_a$ > $\Delta U_c$ and $\Delta U_b$ > $\Delta U_c$ hold, a-b-g fault is specified, whereas, c-g fault is determined if both $\Delta U_c$ > $\Delta U_a$ and $\Delta U_c$ > $\Delta U_b$ hold.

The current-based PS is quite reliable when the short-circuit current source includes not only inverter-interfaced DG units (IIDGs), but also the external grid and/or synchronous DG units. The current-based PS has been defined as primary, since the above is the most common case (especially in looped or meshed networks), while, this kind of PS has proved its efficacy over the years in the field; however, this is not binding. Since this PS is more probable to fail if IIDGs constitute the only short-circuit current source [37], [38] (e.g. in a fully IIDG-based islanded network), in this special case, the primary PS can switch to the voltage-based method (which is expected to perform well [38]). This task is very simple, and could by even performed by the relay vendor in advance, if the relay is meant for networks with the above peculiarity.

*3) Fault direction determination*

After a specific fault type is specified (indicated by a signal *ft* in Fig. 2), the fault direction is determined using a distance element. Although other types of directional elements (e.g. sequence-component-based) could be used for this purpose, a properly applied distance element can be a simple, setting-free and practical way to determine fault direction. After receiving signal *ft*, the distance element calculates the impedance angle $\theta_l$ corresponding to the fault loop $l$ (out of six) involved in the fault (see basic-calculations block in Fig. 2). The impedance angle of a phase- (e.g. a-b) and a ground-fault loop (e.g. a-g) is, respectively, equal to:

$$\theta_{a-b} = \arg\left[\left(\overrightarrow{V_a}-\overrightarrow{V_b}\right)\Big/\left(\overrightarrow{I_{a,s}}-\overrightarrow{I_{b,s}}\right)\right]$$
$$\theta_{a-g} = \arg\left[\overrightarrow{V_a}\Big/\left[\overrightarrow{I_{a,s}}+\overrightarrow{K_0}\cdot\left(3\cdot\overrightarrow{I_{0,s}}\right)\right]\right] \quad (4)$$

where $\overrightarrow{V_a}$, $\overrightarrow{V_b}$, $\overrightarrow{I_{a,s}}$, $\overrightarrow{I_{b,s}}$, are the phase-a voltage, phase-b voltage, superimposed phase-a current and superimposed phase-b current phasors, respectively, measured by the relay, and $\vec{K_0}$ is the zero-sequence compensation factor [35].

Note that superimposed currents are used for $\theta_l$ calculation, to keep the latter uninfluenced by the pre-fault load. If the voltage of the faulted loop is not adequate (e.g. in case of close-in unbalanced faults), the voltage of the healthy phases is used [35]. Also note that setting $\vec{K_0}$ accurately, based on the impedance of the protected line, is important in multi-zone distance protection applications [39], where discrimination between distance zones must be ensured. However, since here the distance element is used for fault direction determination only (i.e. without applying distance zones), absolute $\vec{K_0}$ setting accuracy is not required and a value based on the impedance of a typical MV OH line [35] can be applied by default.

In principle, the calculated faulted-loop impedance lies in the first quadrant of the complex impedance (R-X) plane during a forward fault and in the third quadrant during a reverse fault [35]. However, intermediate infeeds and fault resistance can cause underreach or overreach of the distance element (see Fig. 3a). Since overreach can even lead to angle $\theta_l$ lying in the fourth/second quadrant [14], a forward fault is determined if

$$-90° < \theta_l < 90° \quad (5)$$

whereas, a reverse fault is determined if

$$90° < \theta_l < 270° \quad (6)$$

*4) Trip decision*

If (5) holds (forward fault), a signal *cf* is generated (see Fig. 2) to cancel signal $bf_{i,j}$, blocking the forward operation of $R_{j,i}$. If the respective signal $bf_{j,i}$ from $R_{j,i}$ is not active, $R_{i,j}$ trips $CB_{i,j}$. If (6) holds (reverse fault), a signal *cr* is generated (see Fig. 2) to cancel signal $br_{i,j}$, blocking the reverse operation of the rest $GR_i$-relays. If the blocking signals from the rest $GR_i$-relays (shown as a single signal $br_i$ in Fig. 2) are not active, $R_{i,j}$ sends a signal *sg* to the reverse-fault block (RFB) (see Fig. 2). RFB, described in detail later, discriminates between a reverse fault at bus $B_i$ and a reverse fault in lateral $L_i$, and, in the latter case, ensures coordination between $GR_i$-relays and $P_i$.

Since, after $t_S$ expires, any signal $bf_{j,i}$ (resp. signals $br_i$), sent by $R_{j,i}$ (resp. $GR_i$-relays), must be cancelled by the transmitting relay(s) for a trip to be issued (resp. for a signal *sg* to be sent), the tripping time $t_{TF}$ of $R_{i,j}$ during a forward fault in $L_{i,j}$ (resp. the time $t_{SR}$ to send *sg* during a reverse fault in $B_i/L_i$) will be:

$$\begin{aligned} t_{TF} &= \max\{t_S, t_{SO}\} \\ t_{SR} &= \max\{t_S, t_{S,1}, t_{S,2}, \ldots, t_{S,NR}\} \end{aligned} \quad (7)$$

where $t_{SO}$ is the starting time of $R_{j,i}$ and $t_{S,1}, t_{S,2}, \ldots, t_{S,NR}$ are the starting times of the rest $GR_i$-relays (NR in total). Note that, assuming a fast means, communication latency can be considered negligible (e.g. it equals 4.9 μs/km for fiber optic cable).

### B. 2PH-Fault Protection Element
*1) Starting*

Current unbalance (*iu*) is applied as starting criterion for the 2PH-fault protection element. The most accurate definition of *iu* is [40]:

$$iu = I_2/I_1 \quad (8)$$

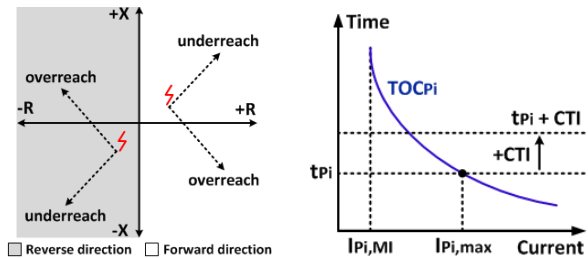

Fig. 3. (a) Applied distance element, (b) 51-element of RFB.

where $I_1$ and $I_2$ is the magnitude of the actual positive- and negative-sequence current, respectively, measured by the relay. *iu* has been chosen here for the starting criterion, as it characterizes 2PH faults and their unbalanced nature. According to [40], the greatest expected *iu* under normal load (non-fault) conditions is 0.3. The latter value is adopted here as the default starting threshold (see Fig. 2), and can be considered robust under various conditions. Given that only a small arc resistance appears during 2PH faults in distribution systems, the above threshold is always expected to be exceeded during 2PH faults, remaining secure under non-fault conditions.

A signal *p* must be also active in order for the 2PH-fault protection element to start, which is generated as long as adequate $3 \cdot I_{0,s}$ is absent (see Fig. 2). In this way, any overlaps between specifying 2PH and SLG faults are avoided, especially when applying $V_2$ vs. $V_1$ in the next step (also see Table I).

The overall starting time is given by (3) in this case too, while, once *iu* exceeds 0.3 and *p* is active, $bf_{i,j}$ and $br_{i,j}$ are generated, similarly to the previous element (see Fig. 2).

*2) Phase selection (PS)*

For PS, $I_{2,s}$ vs. $I_{1,s}$, as well as $V_2$ vs. $V_1$ (if $I_{2,s}$ vs. $I_{1,s}$ fails), described previously, are applied.

The logic of fault direction determination and trip decision (steps 3) and 4), respectively) is similar to that of the previous element. Note that if a reverse fault is determined and any signals $br_i$ are not active, a signal $sp_1$ is sent to RFB (Fig. 2).

### C. 3PH-Fault Protection Element
*1) Starting*

Given that a balanced fault occurs, the following condition must hold so as for the 3PH-fault protection element to start:

$$\left(I_{a,s} \geq 5\%I_n\right) \text{ AND } \left(I_{b,s} \geq 5\%I_n\right) \text{ AND } \left(I_{c,s} \geq 5\%I_n\right) \quad (9)$$

where $I_{a,s}$, $I_{b,s}$ and $I_{c,s}$ is the magnitude of the superimposed phase-a, phase-b and phase-c current, respectively. The above starting criterion is chosen as it characterizes only 3PH faults (i.e. overlaps with other protection elements are avoided). The superimposed phase currents are examined to enhance sensitivity, disengaging the overcurrent threshold from the pre-fault load. The default threshold $5\%I_n$ is chosen, following the same philosophy as that explained for (2). Despite sensitization, protection security is ensured by the proposed protection/communication logic (see Section II). Undesired tripping due to a load change is also avoided, as explained later.

The overall starting time is given by (3). Once (9) asserts, $bf_{i,j}$ and $br_{i,j}$ are generated, as in the previous elements (Fig. 2).

*2) Fault direction determination*

Since 3PH faults are balanced, PS is not applied. Instead, it is directly checked whether all the angles $\theta_l$ of the six fault loops ($\theta_{l,tot}$) satisfy (5) (forward fault) or (6) (reverse fault). Except for indicating fault direction, fulfilling one of these conditions, along with (9), could signify a 3PH fault. If, during close-in 3PH faults, fault-loops voltage is not adequate, the pre-fault (memorized) voltage is used [35] to calculate $\theta_{l,tot}$.

*3) Trip decision*

The trip-decision logic is similar to that of the previous elements. Note that if a reverse fault is determined and any signals $br_i$ are not active, a signal $sp_2$ is sent to RFB (Fig. 2).

Theoretically, the fulfillment of (9) and (5) or (6) (by $\theta_{l,tot}$) could be also due to load rise. However, since load in a segment $L_{i,j}$ is unidirectional, a load rise will not be seen as a forward 3PH fault by one of the two relays protecting $L_{i,j}$. Hence, this relay will not cancel the *bf*-signal sent to its opponent relay (which sees a 3PH fault), avoiding incorrect tripping. In contrast, a load rise could theoretically "mislead" $GR_i$-relays into seeing a reverse 3PH fault. This is dealt with by RFB.

Finally, in the three elements described above, *cf/cr* can be generated not only in the trip-decision step, but also if, in any step, it is realized that an actual fault has not occurred (Fig. 2).

### D. Reverse-Fault Block (RFB)

Once one of the three basic protection elements of a relay $R_{i,j}$ specifies a reverse fault and generates an *s*-signal, the role of RFB (see Fig. 2) is to discriminate between a fault at bus $B_i$ and a fault in lateral $L_i$. In the former case, RFB trips instantaneously, whereas, in the latter case, it introduces a proper time delay to coordinate $R_{i,j}$ with the lateral protection means $P_i$.

Let us explain the basic logic used by RFB to ensure coordination between $R_{i,j}$ and $P_i$. In the first place, the magnitudes of the actual phase currents flowing through $P_i$ ($I_{Pi,a}$, $I_{Pi,b}$, $I_{Pi,c}$) are continuously measured by $R_{i,j}$ through a CT at $P_i$ location. From these currents, the maximum current magnitude $I_{Pi,\max}$ is extracted (see basic-calculations block in Fig. 2). The latter is the current magnitude that would make $P_i$ operate, in case of a fault in $L_i$, and would determine its operating time, based on its time-overcurrent characteristic ($TOC_{Pi}$). To coordinate $R_{i,j}$ with $P_i$, RFB applies an overcurrent (51) element, which maps $I_{Pi,\max}$ onto $TOC_{Pi}$, as shown in Fig. 3b. If $P_i$ is a fuse (resp. an overcurrent relay), its known total-clearing (resp. tripping) characteristic is considered as $TOC_{Pi}$. The latter is uploaded to each $GR_i$-relay and is the only information inserted by the user. Note, however, that this is a simple and effortless task. Using this logic, RFB calculates the melting/tripping time of $P_i$ ($t_{Pi}$), based on $TOC_{Pi}$ and the measured current $I_{Pi,\max}$. Then, it adds a default coordination time interval (*CTI*) to this melting/tripping time, to determine the required time delay of $R_{i,j}$, so that coordination with $P_i$ is ensured (see Fig. 3b). Hence, no matter the fault/system conditions, RFB achieves coordination with $P_i$ by adjusting the time delay of $R_{i,j}$, based on the varying measured $I_{Pi,\max}$ and the default $TOC_{Pi}$ and *CTI*.

The designed internal logic of RFB, concerning both its delayed operation for coordination purposes and its instantaneous operation in any other reverse-fault case, is shown in Fig. 2 and is also described next.

*1) Delayed operation*

First of all, RFB realizes that a fault in $L_i$ has occurred when $I_{Pi,\max}$ starts intersecting $TOC_{Pi}$. This happens when:

$$I_{Pi,\max} \geq I_{Pi,MI} \quad (10)$$

$I_{Pi,MI}$ is the minimum intersecting current of $TOC_{Pi}$ (Fig. 3b).

To ensure that a reverse fault has indeed occurred, *sg*, $sp_1$, or $sp_2$ must also be sent (at $t = t_{SR}$) to initiate the timer of 51-element. This timer expires after $t_{Pi}$, which is calculated as described previously. Considering the *CTI* added for coordination, $R_{i,j}$ will be ultimately ready to trip at:

$$t_{TR} = t_{SR} + t_{Pi} + CTI \quad (11)$$

In this work, a default *CTI* of 0.3 s is considered, which is the most frequent *CTI* value used by protection engineers [1].

If, after $t_{TR}$ expires, (10) still holds, it means that $P_i$ failed to blow/trip, so $R_{i,j}$ trips $CB_{i,j}$, acting as backup protection for lateral $L_i$. Of course, the rest $GR_i$-relays operate similarly.

*2) Instantaneous operation*

If $I_{Pi,\max} < I_{Pi,MI}$ holds, this could mean either that:
i) A fault has not occurred at $B_i$ or $L_i$.
ii) A fault has occurred at $B_i$.
iii) A (ground) fault with considerable resistance, not detectable by $P_i$, has occurred in $L_i$.
iv) A remote fault, not detectable by $P_i$, has occurred at the secondary side of a distribution transformer of $L_i$.

Case-i is excluded if a signal *sg*, $sp_1$, or $sp_2$ is sent. It is also desirable that $GR_i$-relays not trip in case-iv (addressed later).

To recognize case-ii and case-iii faults in a simple way, an undervoltage (27) element is examined only if $I_{Pi,\max} < I_{Pi,MI}$ (see Fig. 2), and is set by default to assert if:

$$(V_a < 0.9 \text{ p.u.}) \text{ OR } (V_b < 0.9 \text{ p.u.}) \text{ OR } (V_c < 0.9 \text{ p.u.}) \quad (12)$$

where $V_a$, $V_b$, $V_c$ are the magnitudes of the $B_i$ phase voltages and 0.9 per unit (p.u.) is the voltage limit indicating an undervoltage disturbance [40]. Two cases are examined:

- If 27-element asserts and *sg*, $sp_1$, or $sp_2$ is generated as well, then (since it is known that $I_{Pi,\max} < I_{Pi,MI}$), a $B_i$-fault is recognized (case-ii) and the relay trips instantaneously.
- If $I_{Pi,\max} < I_{Pi,MI}$ holds, 27-element does not assert, but a signal *sg* is generated, then, a (ground) fault with considerable resistance has expectedly occurred at $B_i$ (case-ii) or $L_i$ (case-iii), so the relay trips again instantaneously.

### E. Protection Reliability Aspects

In the special case where $sp_2$ is due to a load rise, neither (10) nor (12) should hold, so RFB will not trip undesirably.

Assuming distribution transformers which are not double-side grounded (as it is usually the case in many countries), a ground fault in the low-voltage (LV) network (previous case-iv) will not be detected by the ground-fault protection element, due to zero-sequence current absence; hence, undesired RFB trip is avoided. The same also holds in case of load unbalances, given the negligible capacitance of OH distribution lines.

However, even if double-side-grounded distribution transformers are connected to $L_i$, this can be addressed by the vendor/user of the PnP relay, by disabling the part of RFB shown in Fig. 2 into the dashed frame. Hence, the generation of a *sg*-signal alone is no longer enough for a trip to be issued (since now it could theoretically be due to a change in an unbalanced load or a LV ground fault), but either (10) or (12) must also hold. To detect bus faults with high resistance in that case, *sg*-generation in combination with the non-detection of a forward ground fault at $P_i$ location could be used as a signature. Also, only in that case, faults with high resistance in $L_i$ are recognized by RFB as long as (10) holds. Note, however, that a lateral protection means can sense faults with a considerable resistance (as shown later in Section IV), while, it must be borne in mind that $GR_i$-relays only secondarily protect $L_i$, meaning that their actual purpose is to just recognize all the faults that $P_i$ is designed to protect against.

If $GR_i$-relays see a 2PH/3PH fault due to a LV-fault, RFB will not trip, unless (10) holds. Also note that if a LV-fault results in $I_{Pi,\max} < I_{Pi,MI}$, 27-element is not expected to assert.

In the case where one or more DG units are connected to $L_i$, downstream to $P_i$, and the fault current injected by them is strong, (10) could also hold for a $B_i$-fault, "misleading" RFB into applying a time delay, instead of instantaneously tripping. For that reason, RFB is supplemented with the dotted part of Fig. 2, which checks the direction of the fault current flowing through $P_i$. Specifically, based on $B_i$ phase voltages and the superimposed phase currents at $P_i$ ($\overline{I_{Pi,a,s}}$, $\overline{I_{Pi,b,s}}$, $\overline{I_{Pi,c,s}}$), the impedance angle(s) $\theta_{l,Pi}$ ($\theta_{l,tot,Pi}$) of the fault loop(s) indicated by signal $ft$ is (are) calculated by a distance element. Fault direction is specified based on (5) or (6). If a reverse fault (i.e. leading to fault current flowing towards $B_i$) is determined, (10) holds (indicated by a signal $s$) and $sg$, $sp_1$, or $sp_2$ is sent, the above case is recognized, so, instantaneous trip is issued.

It should be also mentioned that the proposed scheme can detect ground faults with a considerable resistance, greater than the resistance dealt with by conventional distribution system protection schemes. This is because, the respective starting function is sensitized as much as possible, by examining the superimposed residual current and considering a typical maximum current sensitivity of commercial relays as the default overcurrent starting threshold. However, there might be cases where high-resistance faults lead to short-circuit currents even lower than the inherent maximum current sensitivity of the relay. In such cases, special methods are required for fault detection, which is an issue requiring dedicated study [41].

Finally, although out of scope, communication operability can be ensured through signal-exchange check or redundancy.

*F. Backup Protection of Main Line Segments and Buses*

For backup protection of main line segments and buses (a lateral $L_i$ is secondarily protected by $GR_i$-relays), a breaker-failure (BF) protection element is designed (see Fig. 2). The BF element ensures that if a CB fails to open, its adjacent CBs will be tripped instead. Specifically, once a relay $R_{i,j}$ issues a trip order, either for main line protection (forward faults), or bus protection (reverse faults), a breaker-failure initiation (BFI) signal is simultaneously generated ($BFI_{fi,j}$ and $BFI_{ri,j}$, respectively), as shown in Fig. 2. Then, the state of $CB_{i,j}$ is checked for a time period $t_{BF}$, compensating for the $CB_{i,j}$ interrupting time. A $t_{BF}$ of 0.24 s is adopted here by default; this corresponds to the maximum typical BF-timer setting according to [42], indicatively considering a system frequency of 50 Hz. If $CB_{i,j}$ remains closed after $t_{BF}$ expires, a trip order is sent to its adjacent CBs. Note that the designed BF element discriminates between a BFI signal generated due to a forward or a reverse fault, so as to trip the proper adjacent CBs ($CB_{fi,j}$ and $CB_{ri,j}$, respectively, in Fig. 2). Once a BFI signal is sent, an instantaneous redundant trip (re-trip) signal is also sent to $CB_{i,j}$, as an attempt to avoid tripping the adjacent CBs.

*G. Economic Evaluation*

It is a fact that the use of communication means as part of the proposed protection scheme requires an investment that is not negligible. However, one should bear in mind that the operation of distribution networks in a looped or meshed configuration is envisioned as a solution to maximize the reliability of future distribution networks, especially in light of increasing DG penetration [9]. For this purpose, the reliability and the efficacy of protection schemes must be, in turn, enhanced. To design advanced protection schemes which satisfy these reliability/efficacy requirements, the use of communication means might be needed. The required investment should be evaluated, taking into account the operational and maintenance benefits gained by the Distribution System Operators (DSOs) and DG producers, due to the efficacy of the protection scheme [43]. For example, the cost of electricity supply interruptions and the outage cost of DG units are critical measures that should be considered in the economic evaluation of the protection scheme [44], [45]. The trend towards using communication means for protecting distribution systems is also reflected on the numerous research papers adopting this philosophy [3]-[11], [13]-[25], [27], [28], [31], [32]. In fact, in [7], [8], communication-based protection schemes are considered for real looped distribution networks. Based on the above, proper motives are expected to arise in the future for DSOs and DG producers, towards the investment in communication-assisted protection schemes. The investment could be possibly shared between DSOs and DG producers, in return for the increased reliability of the distribution system and the avoidance of unnecessary DG disconnection, respectively.

IV. PERFORMANCE EVALUATION

The PnP scheme is mainly tested on the 20 kV, 50 Hz, OH meshed distribution system of Fig. 4, which is designed using realistic data received by the Hellenic Electricity Distribution Network Operator S.A. (HEDNO S.A.), and includes both PVs and synchronous generators (SGs). The system's basic data are shown in Fig. 4, while, a detailed description is included in [14]. Note that the total load of each lateral is shown as a single load, fed by a single distribution transformer, for illustration simplicity. PnP relays are installed at main line segments according to Section II, operating as described previously. Each relay $R_{i,j}$ is accompanied by a circuit breaker $CB_{i,j}$, installed at the same location, while, in this study, it is assumed receiving current and voltage measurements from a local current and voltage transformer with a ratio of 200:5 and 20000:100, respectively. Laterals ($L_i$) are primarily protected by 30-T fuses ($P_i$), commonly applied in Greece, which are assumed existing in the system.

The system of Fig. 4 is modeled with PowerFactory 2018, used for dynamically simulating faults and extracting the voltage/current signals measured at the relays' locations. Then, these signals are processed with MATLAB, where the rest relay elements and functions are modeled and evaluated.

*A. Main Simulation Results*

The relay $R_{2,3}$, protecting $L_{2,3}$ (along with $R_{3,2}$) and $B_2/L_2$ (along with $R_{2,1}$, $R_{2,10}$), is arbitrarily chosen to demonstrate the scheme's performance against a-b-c, a-b, b-c-g and a-g faults at different locations, in GC/ISL system mode. An arc re-sistance ($R_{arc}$) of 2.5 Ω is considered for 3PH/2PH faults, which is a typical maximum $R_{arc}$ in OH MV networks [35]. This is also considered as the minimum fault resistance ($R_{f,min}$) for 2PHG faults. A maximum fault resistance ($R_{f,max}$) of 100 Ω is considered for SLG/2PHG faults. Tables II-V present the critical measurements/calculations of $R_{2,3}$, referring to each algorithm-step, during the above faults at the midpoint of $L_{2,3}$ (forward faults), as well as at buses $B_2$ (reverse bus-faults) and $B_{2.R}$ (reverse lateral-faults). Fault positions are shown in Fig. 4.

The results of Tables II-IV reflect the performance of $R_{2,3}$ until the latter trips (in case of a forward fault) or sends an *s*-signal to RFB (in case of a reverse bus/lateral fault). First of all, based on the $t_S$ values shown (refer to (3)), the starting functions of $R_{2,3}$ always quickly detect the faults and assert.

Then, the fault type (regarding only 2PH/2PHG/SLG faults) is always correctly specified by both the corresponding primary and backup phase selector, as can be realized by the $\varphi$ values of Tables II and III (compare these values to the respective angle ranges of Table I for each fault type), as well as the $\Delta U$ values used to assist fault-type distinction in $V_2$ vs. $V_0$ PS (see Subsection III.A.2). Only in one case in Table II (bolded), the $I_{2,s}$ vs. $I_{1,s}$ PS fails to specify the fault type, which is, however, then correctly specified by the backup $V_2$ vs. $V_0$ PS. Note that the results for the backup PS are shown even in the rest fault cases, just to show its effectiveness.

Fault direction is correctly determined in all the examined fault cases, as the calculated impedance angle(s) ($\theta_l$ in Tables II, III and $\theta_{l,tot}$ in Table IV) corresponding to each fault type examined (see the basic-calculations block of Fig. 2), lie(s) within (-90°, 90°) during forward (main line) faults and (90°, 270°) during reverse (bus/lateral) faults (refer to (5), (6)).

Based on the $t_{TF}$ and $t_{SR}$ values of Tables II-IV, it appears that the relay trips (in case of forward main line faults in Tables II, IV) or sends an *s*-signal to RFB (in case of reverse bus/lateral faults in Tables III, IV) quickly after fault inception. $t_{TF}$ and $t_{SR}$ either coincide with the respective $t_S$ values or they are slightly higher, depending on the time needed for the opponent/adjacent relay/s to process their input signals and cancel the blocking signal initially sent to $R_{2,3}$ (refer to (7)).

Table V shows the critical measurements/calculations of RFB of $R_{2,3}$, during reverse main line bus (bus $B_2$) and lateral (bus $B_{2.R}$) faults. During faults at $B_2$, $R_{2,3}$ measures a current $I_{P2,max}$ flowing through fuse $P_2$, which does not intersect its time-overcurrent characteristic ($TOC_{P2}$), as it is lower than its minimum intersecting current $I_{P2,MI}$ (which equals 74 A, as shown in Fig. 4). Hence, according to Fig. 2, RFB recognizes a non-lateral fault and checks its undervoltage element. As shown in Table V, in case of low-resistance or solid faults at $B_2$, at least one of the three phase voltages drops below 0.9 p.u. (refer to (12)), so, RFB trips instantaneously (as $t_{TR}$ equals the respective $t_{SR}$ of Table III, i.e. the time instant when a *s*-signal is sent to RFB). In case of 100-Ω ground faults at $B_2$, the undervoltage element does not assert; however, as a *sg*-signal has been sent to RFB from the ground-fault protection element, the former recognizes a high-resistance fault and trips instantaneously (as $t_{TR} = t_{SR}$). As for reverse faults at $B_{2.R}$, $R_{2,3}$ always measures a current $I_{P2,max}$ which intersects $TOC_{P2}$ (see Table V). Hence, RFB recognizes a lateral fault and calculates the tripping time $t_{TR}$, so as to ensure coordination with $P_2$. The $t_{TR}$ values of Table V, in case of $B_{2.R}$-faults, are calculated based on the melting time $t_{P2}$ of $P_2$ for the respective measured $I_{P2,max}$ (see $t_{P2}$ values in Table V) and the *CTI* added for coordination (refer to Fig. 2 and (11)). Note that, in case of $B_{2.R}$-faults, we always assume that $P_2$ fails to blow. Just to mention, the high $t_{TR}$ in some cases is due to the high $t_{P2}$.

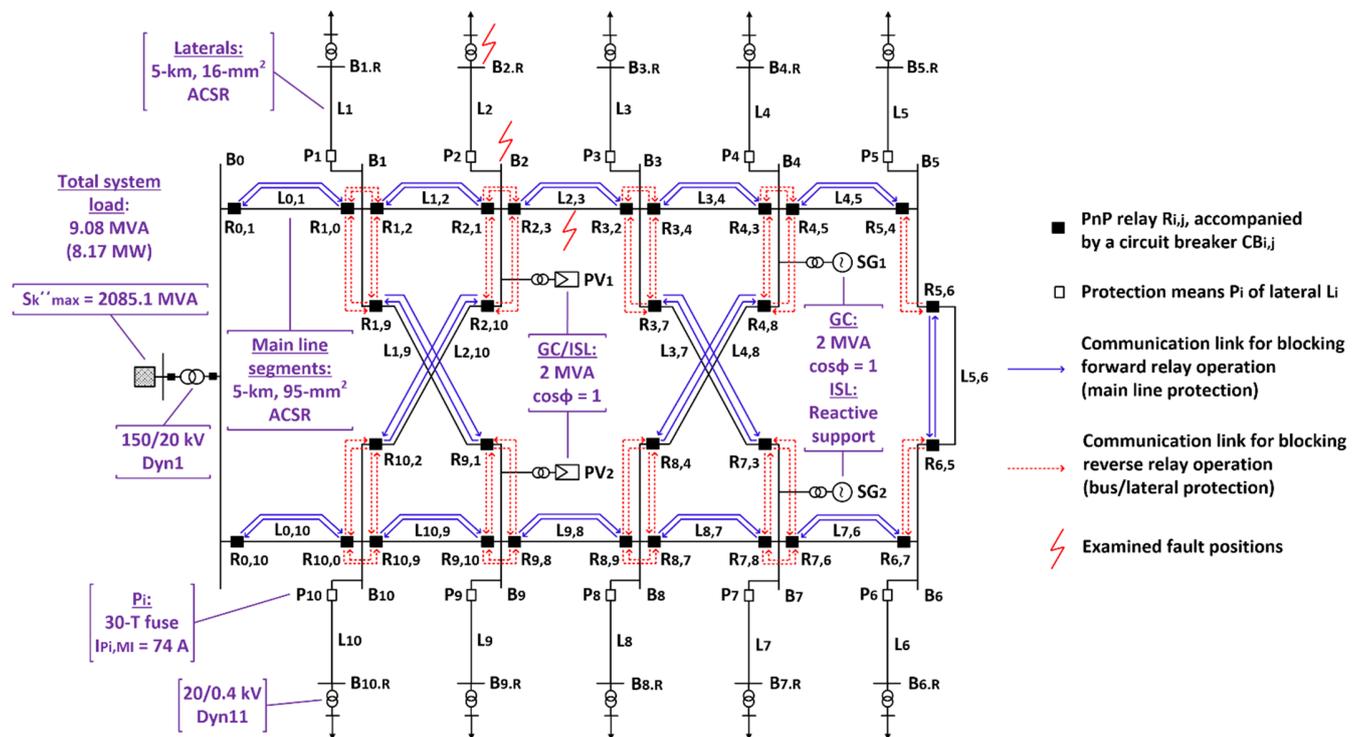

Fig. 4. Test distribution system.

TABLE II
CRITICAL CALCULATIONS OF RELAY $R_{2,3}$ DURING UNBALANCED FAULTS AT THE MIDPOINT OF $L_{2,3}$

| Fault type | $R_{f,min}/R_{f,max}$ (Ω) | Grid-connected mode ||||||||| Islanded mode |||||||||
|---|---|---|---|---|---|---|---|---|---|---|---|---|---|---|---|---|---|---|
| | | $t_S$ (ms) | $\varphi_{Is,21}$ (°) | $\varphi_{Is,20}$ (°) | $\varphi_{V,21}$ (°) | $\varphi_{V,20}$ (°) | $\Delta U_a$ (%) | $\Delta U_b$ (%) | $\Delta U_c$ (%) | $\theta_l$ (°) | $t_{TF}$ (ms) | $t_S$ (ms) | $\varphi_{Is,21}$ (°) | $\varphi_{Is,20}$ (°) | $\varphi_{V,21}$ (°) | $\varphi_{V,20}$ (°) | $\Delta U_a$ (%) | $\Delta U_b$ (%) | $\Delta U_c$ (%) | $\theta_l$ (°) | $t_{TF}$ (ms) |
| a-b | 2.5 | 20.5 | 60.9 | - | 269.8 | - | - | - | - | 20.0 | 20.5 | 22.6 | 61.7 | - | 252.8 | - | - | - | - | 8.2 | 22.6 |
| b-c-g | 2.5/100 | 20.3/22.2 | 186.5/180.2 | 15.0/3.1 | -/- | 5.3/-6.6 | 1.2/1.2 | 43.4/2.0 | 40.9/1.9 | 9.5/-1.5 | 20.3/22.2 | 20.4/22.4 | 178.9/189.0 | -0.1/5.8 | -/- | -15.3/-9.4 | 31.1/2.4 | 69.8/2.7 | 74.2/5.5 | 4.9/0.9 | 20.4/22.4 |
| a-g | 0/100 | 20.2/22.2 | 1.9/-0.6 | 2.2/2.2 | -/- | -7.5/-7.5 | 72.0/2.0 | -3.3[a]/1.1 | 4.8/1.2 | 56.7/-12.5 | 20.2/22.2 | 21.2/22.5 | -0.2/**20.6** | 8.6/8.6 | -/- | -6.5/-6.5 | 90.5/4.4 | 7.7/3.2 | 15.0/0.3 | 57.1/-14.0 | 21.2/22.5 |

[a] Negative voltage drop signifies voltage rise.

TABLE III
CRITICAL CALCULATIONS OF RELAY $R_{2,3}$ DURING UNBALANCED FAULTS AT BUSES $B_2$ AND $B_{2,R}$

| Fault type | Faulted bus | $R_{f,min}/R_{f,max}$ (Ω) | Grid-connected mode ||||||||| Islanded mode |||||||||
|---|---|---|---|---|---|---|---|---|---|---|---|---|---|---|---|---|---|---|---|
| | | | $t_S$ (ms) | $\varphi_{Is,21}$ (°) | $\varphi_{Is,20}$ (°) | $\varphi_{V,21}$ (°) | $\varphi_{V,20}$ (°) | $\Delta U_a$ (%) | $\Delta U_b$ (%) | $\Delta U_c$ (%) | $\theta_l$ (°) | $t_{SR}$ (ms) | $t_S$ (ms) | $\varphi_{Is,21}$ (°) | $\varphi_{Is,20}$ (°) | $\varphi_{V,21}$ (°) | $\varphi_{V,20}$ (°) | $\Delta U_a$ (%) | $\Delta U_b$ (%) | $\Delta U_c$ (%) | $\theta_l$ (°) | $t_{SR}$ (ms) |
| a-b | $B_2$ | 2.5 | 21.2 | 58.1 | - | 272.2 | - | - | - | - | 187.3 | 21.2 | 21.3 | 59.4 | - | 252.3 | - | - | - | - | 187.8 | 21.4 |
| a-b | $B_{2.R}$ | 2.5 | 21.8 | 59.1 | - | 287.4 | - | - | - | - | 199.3 | 21.8 | 21.6 | 58.7 | - | 216.6 | - | - | - | - | 202.5 | 21.6 |
| b-c-g | $B_2$ | 2.5/100 | 20.4/23.6 | 184.0/178.5 | 7.3/-4.0 | -/- | 3.44/-7.9 | -4.6[a]/1.2 | 56.6/2.5 | 41.6/1.8 | 181.0/181.0 | 21.2/30.7 | 21.2/23.6 | 171.9/172.7 | -16.9/-10.6 | -/- | -16.9/-10.6 | 28.7/2.2 | 76.9/3.2 | 72.7/5.6 | 183.7/179.5 | 21.6/31.7 |
| b-c-g | $B_{2.R}$ | 2.5/100 | 21.7/24.2 | 188.0/180.6 | 18.6/-0.8 | -/- | 14.7/-4.6 | -0.7[a]/1.1 | 17.8/2.4 | 19.1/1.8 | 195.2/182.2 | 22.1/31.5 | 21.7/24.2 | 180.0/174.5 | -3.1/-6.9 | -/- | -3.1/-6.9 | 18.7/2.2 | 39.0/3.1 | 46.0/5.6 | 196.7/180.5 | 22.6/32.4 |
| a-g | $B_2$ | 0/100 | 20.3/23.6 | -4.0/0.3 | -4.7/-4.7 | -/- | -8.5/-8.5 | 99.9/2.1 | -10.8[a]/0.8 | 0.8/1.5 | 238.5[b]/164.4 | 20.4/30.7 | 21.4/23.7 | 0.0/-4.3 | -7.3/-7.3 | -/- | -7.3/-7.3 | 100.0/4.6 | 5.6/2.9 | 15.1/2.7 | 245.3[b]/168.8 | 22.0/31.8 |
| a-g | $B_{2.R}$ | 0/100 | 21.4/24.1 | -1.3/0.9 | -4.6/-4.6 | -/- | -8.4/-8.4 | 26.9/2.1 | -3.1[a]/0.8 | 2.8/1.5 | 204.8/166.4 | 21.8/31.4 | 21.7/24.2 | -0.6/-8.3 | -7.2/-7.2 | -/- | -7.2/-7.2 | 47.2/4.0 | 8.6/2.6 | 7.7/2.5 | 201.6/169.6 | 22.6/32.5 |

[a] Negative voltage drop signifies voltage rise.
[b] The voltage of the healthy phases (cross-polarization) has been used to calculate $\theta_{a-g}$, due to inadequate voltage magnitude in the faulted phase.

TABLE IV
CRITICAL CALCULATIONS OF RELAY $R_{2,3}$ DURING 2.5-Ω 3PH FAULTS AT THE MIDPOINT OF $L_{2,3}$ AND BUSES $B_2$ AND $B_{2,R}$

| Fault position | Grid-connected mode ||||||||| Islanded mode |||||||||
|---|---|---|---|---|---|---|---|---|---|---|---|---|---|---|---|---|
| | $t_S$ (ms) | $\theta_{a-b}$ (°) | $\theta_{b-c}$ (°) | $\theta_{c-a}$ (°) | $\theta_{a-g}$ (°) | $\theta_{b-g}$ (°) | $\theta_{c-g}$ (°) | $t_{TF}$ (ms) | $t_{SR}$ (ms) | $t_S$ (ms) | $\theta_{a-b}$ (°) | $\theta_{b-c}$ (°) | $\theta_{c-a}$ (°) | $\theta_{a-g}$ (°) | $\theta_{b-g}$ (°) | $\theta_{c-g}$ (°) | $t_{TF}$ (ms) | $t_{SR}$ (ms) |
| $L_{2,3}$ midpoint | 20.3 | 10.2 | 8.6 | 9.6 | 9.5 | 9.4 | 9.5 | 20.3 | - | 20.1 | 6.9 | 4.4 | 5.1 | 5.1 | 6.6 | 5.0 | 20.2 | - |
| $B_2$ | 20.1 | 182.4 | 181.6 | 181.8 | 181.9 | 182.1 | 181.8 | - | 20.2 | 20.2 | 180.3 | 179.7 | 180.4 | 180.3 | 179.8 | 180.3 | - | 20.2 |
| $B_{2.R}$ | 20.1 | 196.8 | 194.3 | 196.2 | 195.8 | 196.2 | 195.3 | - | 20.2 | 20.2 | 193.6 | 193.2 | 192.6 | 193.3 | 192.6 | 193.4 | - | 20.2 |

TABLE V
CRITICAL MEASUREMENTS/CALCULATIONS OF RFB OF RELAY $R_{2,3}$ DURING FAULTS AT BUSES $B_2$ AND $B_{2,R}$

| Fault type | Faulted bus | $R_{f,min}/R_{f,max}$ (Ω) | Grid-connected mode |||||| Islanded mode ||||||
|---|---|---|---|---|---|---|---|---|---|---|---|---|---|---|
| | | | $I_{P2,max}$ (pri. A) | $V_a$ (% p.u.) | $V_b$ (% p.u.) | $V_c$ (% p.u.) | $t_{P2}$ (ms) | $t_{TR}$ (ms) | $I_{P2,max}$ (pri. A) | $V_a$ (% p.u.) | $V_b$ (% p.u.) | $V_c$ (% p.u.) | $t_{P2}$ (ms) | $t_{TR}$ (ms) |
| a-b-c | $B_2$ | 2.5 | 14.8 | 53.4 | 47.8 | 57.8 | - | 20.2 | 8.4 | 23.1 | 18.7 | 23.0 | - | 20.2 |
| a-b-c | $B_{2.R}$ | 2.5 | 1069.2 | - | - | - | 107.2 | 427.4 | 701.6 | - | - | - | 220.6 | 540.8 |
| a-b | $B_2$ | 2.5 | 26.0 | 72.5 | 35.8 | 100.6 | - | 21.2 | 26.8 | 58.0 | 42.8 | 100.3 | - | 21.4 |
| a-b | $B_{2.R}$ | 2.5 | 968.7 | - | - | - | 127.0 | 448.8 | 562.2 | - | - | - | 334.0 | 655.6 |
| b-c-g | $B_2$ | 2.5/100 | 21.5/25.9 | 106.2/100.6 | 47.5/99.3 | 60.6/100.0 | -/- | 21.2/30.7 | 15.1/26.3 | 66.9/97.2 | 23.4/96.1 | 27.3/94.0 | -/- | 21.6/31.7 |
| b-c-g | $B_{2.R}$ | 2.5/100 | 1124.3/131.0 | -/- | -/- | -/- | 98.3/10180.0 | 420.4/10511.5 | 754.0/124.6 | -/- | -/- | -/- | 193.5/12341.0 | 516.1/12673.4 |
| a-g | $B_2$ | 0/100 | 24.0/26.1 | 0.1/99.7 | 112.7/101.0 | 101.0/100.3 | -/- | 20.4/30.7 | 23.6/26.6 | 0.0/94.2 | 93.1/95.0 | 84.8/98.2 | -/- | 22.0/31.8 |
| a-g | $B_{2.R}$ | 0/100 | 1075.5/129.3 | -/- | -/- | -/- | 105.9/10614.0 | 427.7/10945.4 | 765.7/123.6 | -/- | -/- | -/- | 188.2/12840.0 | 510.8/13172.5 |

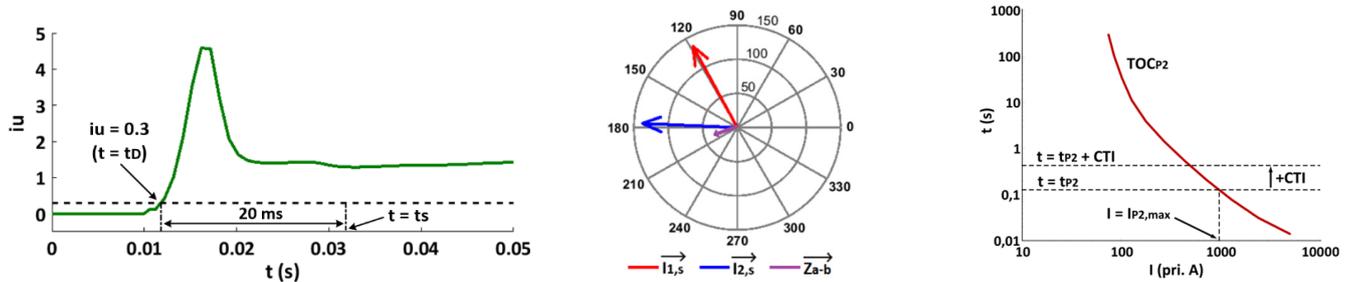

Fig. 5. $R_{2,3}$ calculations during a-b fault at $B_{2.R}$ (GC mode). (a) $iu$ (starting), (b) $\overrightarrow{I_{1,s}}$, $\overrightarrow{I_{2,s}}$ (PS) and $\overrightarrow{Z_{a-b}}$ (loop a-b impedance) at $t = t_S$, (c) Coordination with $P_2$.

It is worth noting that the maximum fault resistance sensed by the ground-fault starting function of a PnP relay can be even greater than the 100-Ω value considered in this study, depending on the fault/system conditions. To provide an example, SLG faults have been simulated at the midpoint of $L_{2,3}$, in order to calculate the maximum fault resistance sensed by $R_{2,3}$ for this specific fault type and fault position. This maximum fault resistance has been found equal to 545 Ω for the GC mode and 529 Ω for the ISL mode of system operation.

Note that 3PH-ground faults would be dealt with similarly to 3PH faults. It is worth mentioning that high-resistance 3PH-ground faults are highly unlikely [16].

In Fig. 5a - Fig. 5c, the representative case of a-b fault at $B_{2.R}$ (GC mode) is chosen, to illustrate the critical operations of $R_{2,3}$. Fault inception is assumed at $t = 0.01$ s. Note that only the primary phase selector is illustrated as part of Fig. 5b.

### B. Additional Simulation Results

To further demonstrate the efficacy of the proposed protection scheme and its independence from specific fault/system conditions, a number of additional simulation scenarios is considered. Note that the default settings of the PnP relays have remained unchanged. In the tables of this subsection, the results regarding only the primary phase selector are given in each case. Moreover, wherever $\theta_{l,tot}$ or $\theta_{l,tot,Pi}$ is given (3PH faults), the impedance angle corresponding to fault loop a-g is shown as a representative value, given that there might by slight divergences between the values of the six fault loops.

*1) Scenario 1: Sudden load change*

In the following, a simulation-based example is given to demonstrate the immunity of the relays protecting a bus/lateral (based on their reverse operation) to a sudden load change.

To simulate a severe load change, we assume that (considering the GC mode of the system) the whole load of lateral $L_2$ is initially disconnected, and is re-connected at $t = 0.01$ s. Fig. 6a and Fig. 6b show the measured phase currents flowing through fuse $P_2$ and the phase voltages measured at bus $B_2$, respectively. As it is apparent in Fig. 6a, all the phase currents are safely below $I_{P2,MI}$, so, (10) does not hold. Hence, a reverse 3PH lateral fault is not mistakenly specified by the relays constituting $GR_2$ (i.e. $R_{2,3}$, $R_{2,1}$ and $R_{2,10}$). Based on the RFB logic described in Subsection III.D, the phase voltages at bus $B_2$ are checked then. As shown in Fig. 6b, these voltages are far from dropping below the default threshold of (12). Therefore, the simulated load change is not mistakenly seen as a reverse 3PH bus fault by the $GR_2$-relays. Obviously, a *sg*-signal is not generated during the load change, so, a ground fault with high resistance is not mistakenly specified either. Moreover, as explained in Subsection III.C.3, a load change would not be of concern as regards the forward operation of the opponent relays protecting a main line segment. To provide an example, during the simulated load change, the opponent relays $R_{2,3}$ and $R_{3,2}$ protecting $L_{2,3}$ (based on their forward operation), calculate an impedance angle ($\theta_{l,tot}$) of approximately 180.4° and 0.8°, respectively; hence, they see this load change in opposite directions, which means that they cannot undesirably trip.

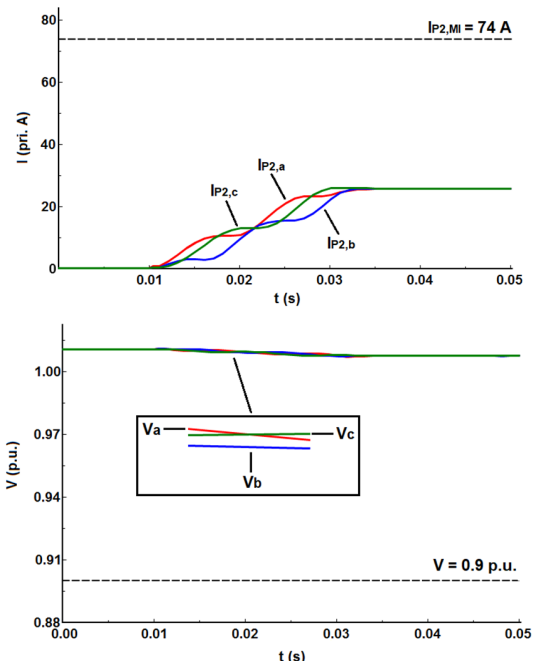

Fig. 6. Scenario 1. (a) Current flowing through $P_2$, (b) Voltage at bus $B_2$.

*2) Scenario 2: Connection of DG to a lateral*

In this scenario, we assume that $SG_1$ is disconnected from main line bus $B_4$ and is connected to the endpoint of lateral $L_2$ (bus $B_{2.R}$). This scenario intends not only to demonstrate the independence of the proposed protection scheme from disconnecting/connecting DG units from/to any network bus, but also to show the ability of a PnP relay to recognize a main line bus fault and trip instantaneously, even if the short-circuit current coming from a DG unit connected to the respective lateral (departing from this bus) intersects the $TOC_{Pi}$ of the lateral protection means (refer to Subsection III.E).

Table VI shows the critical calculations of relay $R_{2,3}$ during faults at bus $B_2$, assuming all the fault-type, fault-resistance and operation mode (O/M) cases considered in the previous subsection. According to the results, the three basic protection elements of the relay always quickly detect the fault, correctly specify the fault type and the fault direction (reverse) and send an *s*-signal to the RFB of the relay, after $t = t_{SR}$ (i.e. after all the relays of $GR_2$ cancel their blocking signals).

Table VII shows the critical measurements/calculations of RFB. In the vast majority of the simulated cases, the measured $I_{P2,max}$ intersects $TOC_{P2}$ (as $I_{P2,max} > 74$ A), due to the short-circuit current contribution of $SG_1$, connected to $B_{2.R}$. This fact "misleads" RFB into specifying a lateral fault instead of a main line bus fault (note that the undervoltage element is ignored in these cases). However, this issue is addressed as explained in Subsection III.E (i.e. through the dotted part of Fig. 2). In particular, the direction of $I_{P2,max}$ is specified as reverse in all cases (as it is derived from the $\theta_{l,P2}/\theta_{l,tot,P2}$ values of Table VII), which indicates a fault at $B_2$. Subsequently, a trip order with no intentional time delay is issued (see the $t_{TR}$ values of Table VII), instead of mistakenly tripping after the respective $t_{P2}$ (also shown in Table VII) and *CTI* elapses.

TABLE VI
CRITICAL CALCULATIONS OF RELAY $R_{2,3}$ DURING FAULTS AT BUS $B_2$ (SCENARIO 2)

| O/M | Fault type | $R_{f,min}/R_{f,max}$ (Ω) | $t_S$ (ms) | $\varphi_{Is,21}$ (°) | $\varphi_{Is,20}$ (°) | $\theta_l/\theta_{l,tot}$ (°) | $t_{SR}$ (ms) |
|---|---|---|---|---|---|---|---|
| GC | a-b-c | 2.5 | 22.2 | - | - | 181.6 | 22.2 |
| GC | a-b | 2.5 | 21.6 | 57.0 | - | 183.1 | 21.6 |
| GC | b-c-g | 2.5/100 | 20.5/29.0 | 178.6/177.1 | 4.2/1.4 | 182.6/182.6 | 21.5/32.6 |
| GC | a-g | 0/100 | 20.4/29.0 | -3.0/-3.0 | 1.3/1.3 | 241.6/170.4 | 21.1/32.6 |
| ISL | a-b-c | 2.5 | 22.7 | - | - | 184.5 | 23.6 |
| ISL | a-b | 2.5 | 21.6 | 58.2 | - | 185.4 | 22.5 |
| ISL | b-c-g | 2.5/100 | 22.1/29.4 | 167.1/176.2 | 27.4/-6.8 | 185.0/185.3 | 22.3/34.1 |
| ISL | a-g | 0/100 | 21.8/29.7 | -1.8/-1.3 | -2.4/-2.4 | 248.1/173.3 | 22.3/34.4 |

TABLE VII
CRITICAL MEASUREMENTS/CALCULATIONS OF RFB OF RELAY $R_{2,3}$ DURING FAULTS AT BUS $B_2$ (SCENARIO 2)

| O/M | Fault type | $R_{f,min}/R_{f,max}$ (Ω) | $I_{P2,max}$ (pri. A) | $V_a$ (% p.u.) | $V_b$ (% p.u.) | $V_c$ (% p.u.) | $t_{P2}$ (ms) | $\theta_{l,P2}/\theta_{l,tot,P2}$ (°) | $t_{TR}$ (ms) |
|---|---|---|---|---|---|---|---|---|---|
| GC | a-b-c | 2.5 | 335.2 | - | - | - | 917.8 | 171.1 | 22.2 |
| GC | a-b | 2.5 | 339.5 | - | - | - | 893.8 | 176.5 | 21.6 |
| GC | b-c-g | 2.5/100 | 394.7/49.9 | -/100.7 | -/99.5 | -/99.8 | 658.0/- | 174.8/174.7 | 21.5/32.6 |
| GC | a-g | 0/100 | 666.6/52.6 | -/99.7 | -/100.9 | -/100.6 | 242.6/- | 218.0/146.9 | 21.1/32.6 |
| ISL | a-b-c | 2.5 | 530.8 | - | - | - | 371.2 | 171.3 | 23.6 |
| ISL | a-b | 2.5 | 428.7 | - | - | - | 555.9 | 172.4 | 22.5 |
| ISL | b-c-g | 2.5/100 | 553.9/95.3 | -/- | -/- | -/- | 343.0/49552.2 | 172.3/171.2 | 22.3/34.1 |
| ISL | a-g | 0/100 | 614.9/90.0 | -/- | -/- | -/- | 282.0/69771.3 | 229.3/153.6 | 22.3/34.4 |

*3) Scenario 3: Loss of network-part and looped configuration*

As part of this scenario, we suppose that $R_{3,2}$, $R_{3,7}$ and $R_{3,4}$ have tripped (e.g. after a fault at $B_3/L_3$) and the respective CBs are open. This fact leads to a loss of short-circuit contribution through both line segments $L_{3,4}$ and $L_{3,7}$, in case of a fault at $L_{2,3}$. Actually, in this case, the relay $R_{2,3}$ protects a radial main line segment, as a fault in $L_{2,3}$ is fed from only one direction.

To show the immunity of a PnP relay (e.g. $R_{2,3}$) to the loss of a network-part, even if the latter leads to loss of the looped configuration, all the fault cases of the previous subsection are simulated at the midpoint of $L_{2,3}$, assuming the above scenario. The simulation results are shown in Table VIII. As it is apparent, $R_{2,3}$ operates quickly and correctly in all the simulated fault cases, sending an immediate trip order to $CB_{2,3}$, so as to isolate $L_{2,3}$. Note that, since the opponent relay $R_{3,2}$ does not initiate its operation for any of the simulated faults (as it does not see any current), it never sends a blocking signal to $R_{2,3}$.

*4) Scenario 4: Loss of DG*

In order to show the independence of the PnP protection scheme from any DG loss, we assume a marginal (worst-case) scenario, where all the DG units of the test distribution system are disconnected. Since no DG is present, only the GC mode of the system is examined. Table IX shows the critical calculations of relay $R_{2,3}$, during faults at the midpoint of $L_{2,3}$, considering the above system case. It appears that the relay always operates quickly and correctly, despite the mass DG loss, further proving the scheme's immunity to DG intermittence.

TABLE VIII
CRITICAL CALCULATIONS OF RELAY $R_{2,3}$ DURING FAULTS AT THE MIDPOINT OF $L_{2,3}$ (SCENARIO 3)

| O/M | Fault type | $R_{f,min}/R_{f,max}$ (Ω) | $t_S$ (ms) | $\varphi_{Is,21}$ (°) | $\varphi_{Is,20}$ (°) | $\theta_l/\theta_{l,tot}$ (°) | $t_{TF}$ (ms) |
|---|---|---|---|---|---|---|---|
| GC | a-b-c | 2.5 | 21.3 | - | - | 15.4 | 21.3 |
| GC | a-b | 2.5 | 20.2 | 60.0 | - | 25.0 | 20.2 |
| GC | b-c-g | 2.5/100 | 20.2/21.4 | 187.6/181.5 | 21.4/3.1 | 15.4/0.5 | 20.2/21.4 |
| GC | a-g | 0/100 | 20.2/21.3 | 0.0/0.0 | 0.0/0.0 | 56.8/-12.5 | 20.2/21.3 |
| ISL | a-b-c | 2.5 | 21.6 | - | - | 15.4 | 21.6 |
| ISL | a-b | 2.5 | 20.2 | 59.9 | - | 25.0 | 20.2 |
| ISL | b-c-g | 2.5/100 | 20.3/21.5 | 178.9/178.5 | -1.7/-2.8 | 15.4/0.4 | 20.3/21.5 |
| ISL | a-g | 0/100 | 20.4/21.5 | 0.0/-0.3 | 0.0/0.0 | 56.8/-12.6 | 20.4/21.5 |

TABLE IX
CRITICAL CALCULATIONS OF RELAY $R_{2,3}$ DURING FAULTS AT THE MIDPOINT OF $L_{2,3}$ (SCENARIO 4)

| Fault type | $R_{f,min}/R_{f,max}$ (Ω) | $t_S$ (ms) | $\varphi_{Is,21}$ (°) | $\varphi_{Is,20}$ (°) | $\theta_l/\theta_{l,tot}$ (°) | $t_{TF}$ (ms) |
|---|---|---|---|---|---|---|
| a-b-c | 2.5 | 21.3 | - | - | 12.2 | 22.2 |
| a-b | 2.5 | 21.7 | 60.0 | - | 20.8 | 21.9 |
| b-c-g | 2.5/100 | 21.2/22.1 | 176.6/181.0 | -28.8/2.4 | 12.2/0.4 | 22.1/29.3 |
| a-g | 0/100 | 20.3/21.9 | 0.0/0.0 | 0.0/0.0 | 56.8/-12.8 | 21.5/25.4 |

*5) Scenario 5: IIDG-based islanded network*

A well-known protection issue when the network is IIDG-dominated, especially if it operates as an island, is the decreased relay sensitivity due to the low short-circuit current contribution of IIDGs. As also explained in Subsection III.E for resistive faults, the proposed PnP relay can detect faults characterized by a quite low short-circuit current, as its starting functions are desensitized as much as possible, in compliance with the capabilities of commercial relays. To consider an extreme case, we assume that the system of Fig. 4 can operate as a fully (100%) IIDG-based island. To this end, $SG_1$ and $SG_2$ are assumed being disconnected, while, two PV units of the same MW rating are instead connected to the respective buses. It is also assumed that all the PV units provide reactive support. Table X includes the critical calculations of relay $R_{2,3}$, during faults at the midpoint of $L_{2,3}$. According to the relevant explanation of Subsection III.A.2, in this special case, we rely on the voltage-based PS. Although the measured short-circuit currents are relatively low in Scenario 5, the starting quantities always quickly exceed the respective starting thresholds. After initiation, the relay always specifies the fault type and the fault direction correctly, issuing an instantaneous trip order.

TABLE X
CRITICAL CALCULATIONS OF RELAY $R_{2,3}$ DURING FAULTS AT THE MIDPOINT OF $L_{2,3}$ (SCENARIO 5)

| Fault type | $R_{f,\min}/R_{f,\max}$ (Ω) | $t_S$ (ms) | $\varphi_{V,21}$ (°) | $\varphi_{V,20}$ (°) | $\Delta U_a$ (%) | $\Delta U_b$ (%) | $\Delta U_c$ (%) | $\theta_i/\theta_{l,tot}$ (°) | $t_{TF}$ (ms) |
|---|---|---|---|---|---|---|---|---|---|
| a-b-c | 2.5 | 21.4 | - | - | - | - | - | 9.6 | 21.4 |
| a-b | 2.5 | 24.9 | 240.7 | - | - | - | - | 15.9 | 24.9 |
| b-c-g | 2.5/ 100 | 27.3/ 22.4 | -/ - | -23.8/ -63.4 | 63.2/ -2.1 | 87.3/ 10.3 | 89.5/ 15.3 | 9.6/ 5.6 | 27.3/ 22.9 |
| a-g | 0/ 100 | 23.1/ 22.4 | -/ - | -58.7/ -58.7 | 96.1/ 15.4 | -7.5/ 0.1 | 6.3/ -5.4 | 56.7/ -9.4 | 23.4/ 22.9 |

*6) Scenario 6: Application to the IEEE 14-bus test system*

To further validate the independence of the proposed PnP protection scheme from specific system conditions, the scheme is applied to the 33-kV (distribution) part of the IEEE 14-bus test system [46], shown in Fig. 7. This part is connected to a 132-kV system, through the transformers feeding buses $B_6$ and $B_9$ of Fig. 7. We assume that the 33-kV system operates in the ISL mode when it is disconnected from these transformers and is fed solely from the synchronous generator SG. Although SG injects only reactive power during the GC mode, it also assumed injecting active power to support the ISL operation. The proposed protection scheme can protect the line segments and buses of the system of Fig. 7 in the same logic as that described previously throughout the paper. In the present example, we focus on the relay $R_{12,6}$ of Fig. 7, protecting line segment $L_{6,12}$ (along with its opponent relay $R_{6,12}$).

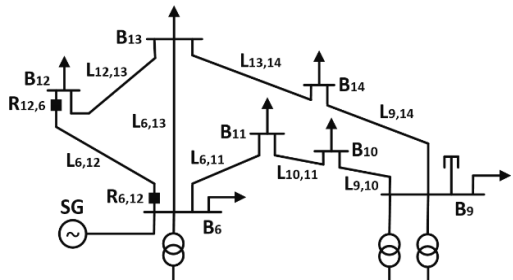

Fig. 7. 33-kV part of the IEEE 14-bus test system.

TABLE XI
CRITICAL CALCULATIONS OF RELAY $R_{12,6}$ DURING FAULTS AT THE MIDPOINT OF $L_{6,12}$ IN THE IEEE 14-BUS TEST SYSTEM (SCENARIO 6)

| O/M | Fault type | $R_{f,\min}/R_{f,\max}$ (Ω) | $t_S$ (ms) | $\varphi_{Is,21}$ (°) | $\varphi_{Is,20}$ (°) | $\theta_i/\theta_{l,tot}$ (°) | $t_{TF}$ (ms) |
|---|---|---|---|---|---|---|---|
| GC | a-b-c | 2.5 | 18.0 | - | - | -3.1 | 18.0 |
|  | a-b | 2.5 | 19.0 | 59.9 | - | 3.4 | 19.0 |
|  | b-c-g | 2.5/ 100 | 16.9/ 23.2 | 171.6/ 180.1 | -14.9/ 0.1 | -3.1/ -9.8 | 16.9/ 23.2 |
|  | a-g | 0/ 100 | 16.9/ 23.2 | 0.0/ 0.8 | 0.7/ 0.7 | 51.3/ -22.7 | 16.9/ 23.2 |
| ISL | a-b-c | 2.5 | 18.0 | - | - | -2.8 | 18.0 |
|  | a-b | 2.5 | 22.0 | 60.0 | - | 3.2 | 22.0 |
|  | b-c-g | 2.5/ 100 | 17.0/ 23.3 | 173.8/ 179.7 | -11.4/ -0.6 | -2.8/ -9.3 | 17.0/ 23.3 |
|  | a-g | 0/ 100 | 17.0/ 23.3 | 0.0/ 0.0 | 0.0/ 0.0 | 50.7/ -22.9 | 17.0/ 23.3 |

Table XI shows the critical calculations of relay $R_{12,6}$, during faults at the midpoint of $L_{6,12}$, in both the GC and the ISL mode of system operation. It appears that the relay operates quickly and correctly in all the simulated fault cases. Just to mention, the opponent relay $R_{6,12}$ also operates correctly in all cases and cancels its blocking signal, letting $R_{12,6}$ trip at $t = t_{TF}$ (as respectively performed by $R_{12,6}$). Note that the starting times ($t_S$) of Table XI result from (3), bearing in mind that a 60-Hz system is examined in this scenario.

V. COMPARISON WITH CURRENT DIFFERENTIAL PROTECTION

The main drawbacks of a differential scheme compared to the proposed PnP scheme are practical, and regard the need for properly setting the differential relays due to sensitivity, security and coordination purposes (unlike the PnP relays), as well as the need for separate schemes/relays to protect a main line segment and its adjacent bus/lateral (also refer to Section I).

However, it would be interesting to also compare the two protection schemes in terms of performance. For this purpose, a differential scheme is assumed protecting each main line segment $L_{i,j}$, while, a differential scheme is also considered protecting each bus/lateral $B_i/L_i$ of the system of Fig. 4. The pickup differential current ($I_{d,p}$) is set equal to 5%$I_n$ for main line differential relays, i.e. as the default threshold of (2) and (9). Assuming a uniform CT ratio of 200:5 (as before), 5%$I_n$ equals 10 pri. A. As for the differential relays protecting buses/laterals, $I_{d,p}$ is set greater than the differential current measured during normal load conditions. To this end, the maximum load current flowing through each lateral and any infeed load currents from DG units are taken into account. A safety factor of 1.2 [11] is also considered in the latter cases.

In this study, faults of all types have been simulated at the midpoint of all segments $L_{i,j}$, at all main line buses $B_i$ and at the endpoint of all laterals $L_i$. The same fault resistance cases as before have been considered, while, each fault has been simulated in both the GC and the ISL mode of the system.

As for the main line segments, both the differential and the PnP scheme have detected and cleared all the faults. Regarding bus/lateral faults, the differential scheme has displayed similar performance in the vast majority of fault cases; however, it has failed to detect and clear most of the 100-Ω ground faults at buses $B_4$ and $B_{4,R}$, as well as $B_7$ and $B_{7,R}$. On the other hand, the PnP scheme have detected and cleared all the faults.

Due to the great amount of results, only those concerning the above critical faults at $B_4$ and $B_{4,R}$ are indicatively provided in Table XII. This table includes the differential current measured by the differential scheme protecting $B_4/L_4$. The values in bold are lower than the corresponding $I_{d,p}$ setting (shown in Table XII), leading to differential protection failure. The main reason for that is the need to desensitize the $I_{d,p}$ setting, due to the connection of $SG_1$ to $B_4$. Exactly the same situation appears during 100-Ω ground faults at $B_7$ and $B_{7,R}$.

Table XII also includes the starting quantities measured by the $GR_4$ PnP relays in the same fault cases. $3 \cdot I_{0,s,4,3}$, $3 \cdot I_{0,s,4,8}$ and $3 \cdot I_{0,s,4,5}$ is the superimposed residual current measured by $R_{4,3}$, $R_{4,8}$ and $R_{4,5}$, respectively. As can be seen, sensitivity issues appear in this case as well, as $R_{4,5}$ fails to detect the faults and start its operation (see the bolded values). In this regard, $R_{4,5}$ does not transmit a blocking signal either. However, the use of the redundant trip (inter-trip) order (see Section II), renders the PnP scheme able to clear a fault, on condition that at least one of the assigned relays senses the fault. Hence, in all the cases of Table XII, $R_{4,3}$ and $R_{4,8}$ detect the fault, tripping not only their own *CB*, but also the *CB* of the rest $GR_4$-relays (including that of $R_{4,5}$). Exactly the same situation appears during 100-Ω ground faults at $B_7$ and $B_{7.R}$. Note that in all the cases of this study, after initiation, the PnP scheme always correctly specifies the fault type and the fault direction.

Just to mention, it would be even easier for the PnP relays protecting $B_4/L_4$ and $B_7/L_7$ to detect the simulated 100-Ω ground faults, if a CT with a $I_n$ lower than 200 pri. A was considered for each relay. This would be permissible, based on the maximum load of the line segments where these relays are installed. However, the 200:5 CT ratio has been universally considered, for the sake of uniformity. On the other hand, a different CT ratio would not affect (sensitize) the $I_{p,d}$ setting of the differential scheme protecting $B_4/L_4$ and $B_7/L_7$.

TABLE XII
PICKUP/STARTING QUANTITIES MEASURED BY THE DIFFERENTIAL AND THE PLUG-AND-PLAY SCHEME DURING 100-Ω GROUND FAULTS AT $B_4/B_{4.R}$

|  | Pickup/starting quantity shown | Fault type | Faulted bus | GC mode | ISL mode |
|---|---|---|---|---|---|
| Differential protection scheme | Differential current (pri. A) ($I_{d,p}$ = 76 pri. A) | b-c-g | $B_4$ | 80.4 | **68.8** |
|  |  |  | $B_{4.R}$ | 78.2 | **64.2** |
|  |  | a-g | $B_4$ | **62.2** | **55.2** |
|  |  |  | $B_{4.R}$ | **54.6** | **51.2** |
| PnP protection scheme | $3 \cdot I_{0,s,4,3}$ (pri. A) $3 \cdot I_{0,s,4,8}$ (pri. A) $3 \cdot I_{0,s,4,5}$ (pri. A) (default starting threshold = 10 pri. A) | b-c-g | $B_4$ | 11.7 11.7 **6.2** | 11.5 11.5 **6.1** |
|  |  |  | $B_{4.R}$ | 10.6 10.6 **5.6** | 10.3 10.3 **5.5** |
|  |  | a-g | $B_4$ | 11.6 11.6 **6.2** | 11.2 11.2 **6.1** |
|  |  |  | $B_{4.R}$ | 10.5 10.5 **5.6** | 10.2 10.2 **5.5** |

## VI. CONCLUSION

To provide a reliable protection solution for future OH looped/meshed distribution systems with DG, eliminating protection design complexity, this paper proposes a protection scheme, which relies on PnP, communication-assisted, numerical relays. The PnP relays are beforehand designed so that they do not require user-defined settings, being unaffected by system changes and independent of a particular network. As such, coordination of main line relays with each other as well as coordination between main line relays and lateral protection means is guaranteed, without needing a coordination study. Only the upload of existing lateral protection means' characteristics to the relays is required, which is, however, a simple and effortless task. Replacement of existing lateral protection means is also not needed. The latter fact, as well as the employment of existing relay capabilities, enhance the scheme's applicability. The simulation results are very promising, as the scheme proves effective against a variety of different fault and system conditions. Hardware experiments, to further validate the proposed scheme, are still pending. It is the authors' goal to conduct such experiments as part of future work.

## VII. REFERENCES

[1] J. L. Blackburn and T. J. Domin, *Protective relaying: Principles and applications*. Boca Raton, FL, USA: CRC Press, 2014.
[2] E. Dehghanpour, H. K. Karegar, R. Kheirollahi, and T. Soleymani, "Optimal coordination of directional overcurrent relays in microgrids by using cuckoo-linear optimization algorithm and fault current limiter," *IEEE Trans. Smart Grid*, vol. 9, no. 2, pp. 1365-1375, Mar. 2018.
[3] H. M. Sharaf, H. H. Zeineldin, and E. El-Saadany, "Protection coordination for microgrids with grid-connected and islanded capabilities using communication assisted dual setting directional overcurrent relays," *IEEE Trans. Smart Grid*, vol. 9, no. 1, pp. 143-151, Jan. 2018.
[4] A. Yazdaninejadi, S. Golshannavaz, D. Nazarpour, S. Teimourzadeh, and F. Aminifar, "Dual-setting directional overcurrent relays for protecting automated distribution networks," *IEEE Trans. Ind. Informat.*, vol. 15, no. 2, pp. 730-740, Feb. 2019.
[5] V. C. Nikolaidis, E. Papanikolaou, and A. S. Safigianni, "A communication-assisted overcurrent protection scheme for radial distribution systems with distributed generation," *IEEE Trans. Smart Grid*, vol. 7, no. 1, pp. 114-123, Jan. 2016.
[6] L. Che, M. E. Khodayar, and M. Shahidehpour, "Adaptive protection system for microgrids: Protection practices of a functional microgrid system," *IEEE Electrific. Mag.*, vol. 2, no. 1, pp. 66-80, Mar. 2014.
[7] J. R. Fairman, K. Zimmerman, J. W. Gregory, and J. K. Niemira, "International drive distribution automation and protection," presented at the 27th Annu. Western Protective Relay Conf., Spokane, WA, USA, 2000.
[8] S. Lauria, A. Codino, and R. Calone, "Protection system studies for ENEL Distribuzione's MV loop lines," in *Proc. PowerTech*, Eindhoven, Netherlands, 2015, pp. 1-6.
[9] E. Sortomme, S. S. Venkata, and J. Mitra, "Microgrid protection using communication-assisted digital relays," *IEEE Trans. Power Del.*, vol. 25, no. 4, pp. 2789-2796, Oct. 2010.
[10] X. Liu, M. Shahidehpour, Z. Li, X. Liu, Y. Cao, and W. Tian, "Protection scheme for loop-based microgrids," *IEEE Trans. Smart Grid*, vol. 8, no. 3, pp. 1340-1349, May 2017.
[11] T. S. Aghdam, H. K. Karegar, and H. H. Zeineldin, "Variable tripping time differential protection for microgrids considering DG stability," *IEEE Trans. Smart Grid*, vol. 10, no. 3, pp. 2407-2415, May 2019.
[12] G. Ziegler, *Numerical Differential Protection: Principles and Applications*. Erlangen, Germany: Publicis Publishing, 2012.
[13] H. F. Habib, T. Youssef, M. H. Cintuglu, and O. Mohammed, "Multi-agent-based technique for fault location, isolation and service restoration," *IEEE. Trans. Ind. Appl.*, vol. 53, no. 3, pp. 1841-1851, May 2017.
[14] A. M. Tsimtsios, G. N. Korres, and V. C. Nikolaidis, "A pilot-based distance protection scheme for meshed distribution systems with distributed generation," *Int. J. Elect. Power Energy Syst.*, vol. 105, pp. 454-469, Feb. 2019.
[15] M. Elkhatib, A. Ellis, M. Biswal, S. Brahma, and S. Ranade, "Protection of renewable-dominated microgrids: Challenges and potential solutions," Sandia National Laboratories, Albuquerque, NM, USA, Rep. SAND2016-11210, Nov. 2016.
[16] M. A. Zamani, A. Yazdani, and T. S. Sidhu, "A communication-assisted protection strategy for inverter-based medium-voltage microgrids," *IEEE Trans. Smart Grid*, vol. 3, no. 4, pp. 2088-2099, Dec. 2012.
[17] C. Yuan, K. Lai, M. S. Illindala, M. A. Haj-ahmed, and A. S. Khalsa, "Multilayered protection strategy for developing community microgrids in village distribution systems," *IEEE Trans. Power Del.*, vol. 32, no. 1, pp. 495-503, Feb. 2017.
[18] V. A. Papaspiliotopoulos, G. N. Korres, V. A. Kleftakis, and N. D. Hatziargyriou, "Hardware-in-the-loop design and optimal setting of adaptive protection schemes for distribution systems with distributed generation," *IEEE Trans. Power Del.*, vol. 32, no. 1, pp. 393-400, Feb. 2017.
[19] E. Purwar, D. N. Vishwakarma, and S. P. Singh, "A novel constraints reduction based optimal relay coordination method considering variable operational status of distribution system with DGs," *IEEE Trans. Smart Grid*, vol. 10, no. 1, pp. 889-898, Jan. 2019.
[20] Z. Liu, H. K. Høidalen, and M. M. Saha, "An intelligent coordinated protection and control strategy for distribution network with wind generation integration," *CSEE J. Power Energy Syst.*, vol. 2, no. 4, pp. 23-30, Dec. 2016.

[21] A. Abbasi, H. K. Karegar, and T. S. Aghdam, "Adaptive protection coordination with setting groups allocation," *Int. J. Renewable Energy Res.*, vol. 9, no. 2, pp. 795-803, Jun. 2019.

[22] H. Muda and P. Jena, "Sequence currents based adaptive protection approach for DNs with distributed energy resources," *IET Gen., Transm. Distrib.*, vol. 11, no. 1, pp. 154-165, May 2017.

[23] Z. Liu, C. Su, H. K. Høidalen, and Z. Chen, "A Multiagent system-based protection and control scheme for distribution system with distributed-generation integration," *IEEE Trans. Power Del.*, vol. 32, no. 1, pp. 536-545, Feb. 2017.

[24] S. A. Hosseini, A. Nasiri, and S. H. H. Sadeghi, "A decentralized adaptive scheme for protection coordination of microgrids based on team working of agents," in *Proc. 7th Int. Conf. Renewable Energy Res. Appl.*, Paris, France, 2018, pp. 1315-1320.

[25] M. Y. Shih, A. Conde, Z. Leonowicz, and L. Martirano, "An adaptive overcurrent coordination scheme to improve relay sensitivity and overcome drawbacks due to distributed generation in smart grids," *IEEE Trans. Ind. Appl.*, vol. 53, no. 6, pp. 5217-5228, Nov.-Dec. 2017.

[26] D. P. Mishra, S. R. Samantaray, and G. Joos, "A combined wavelet and data-mining based intelligent protection scheme for microgrid," *IEEE Trans. Smart Grid*, vol. 7, no. 5, pp. 2295-2304, Sep. 2016.

[27] S. Kar, S. R. Samantaray, and M. D. Zadeh, "Data-mining model based intelligent differential microgrid protection scheme," *IEEE Syst. J.*, vol. 11, no. 2, pp. 1161-1169, Jun. 2017.

[28] M. Mishra and P. K. Rout, "Detection and classification of micro-grid faults based on HHT and machine learning techniques," *IET Gen., Transm. Distrib.*, vol. 12, no. 2, pp. 388-397, Jan. 2018.

[29] J. J. Q. Yu, Y. Hou, A. Y. S. Lam, V. O. K. Li, "Intelligent fault detection scheme for microgrids with wavelet-based deep neural networks," *IEEE Trans. Smart Grid*, vol. 10, no. 2, pp. 1694-1703, Mar. 2019.

[30] Y. Liu, A. P. Meliopoulos, L. Sun, and S. Choi, "Protection and control of microgrids using dynamic state estimation," *Prot. Control Mod. Power Syst.*, vol. 3, no. 1, pp. 1-13, Dec. 2018.

[31] S. B. A. Bukhari, R. Haider, M. S. U. Zaman, Y. Oh, G. Cho, C. Kim, "An interval type-2 fuzzy logic based strategy for microgrid protection," *Int. J. Elect. Power Energy Syst.*, vol. 98, pp. 209-218, Jun. 2018.

[32] Z. Zhang, B. Xu, P. Crossley, and L. Li, "Positive-sequence-fault-component-based blocking pilot protection for closed-loop distribution network with underground cable," *Int. J. Elect. Power Energy Syst.*, vol. 94, pp. 57-66, Jan. 2018.

[33] G. Benmouyal and J. Roberts, "Superimposed quantities: Their true nature and application in relays," presented at the 26th Annu. Western Protective Relay Conf., Spokane, WA, USA, Oct. 26-28, 1999.

[34] *D60 Line Distance Relay*, GE, Markham, ON, Canada, 2009. [Online]. Available: https://www.gegridsolutions.com/products/manuals/d60/d60man-f5.pdf

[35] G. Ziegler, *Numerical Distance Protection: Principles and Applications*. Erlangen, Germany: Publicis Publishing, 2011.

[36] B. Kasztenny, B. Cambell, and J. Mazereeuw, "Phase selection for single-pole tripping–Weak infeed conditions and cross-country faults," presented at the 27th Annu. Western Protective Relay Conf., Spokane, WA, USA, Oct. 24-26, 2000.

[37] M. A. Azzouz, A. Hooshyar, and E. F. El-Saadany, "Resilience enhancement of microgrids with inverter-interfaced DGs by enabling faulty phase selection," *IEEE Trans. Smart Grid*, vol. 9, no. 6, pp. 6578-6589, Nov. 2018.

[38] A. Hooshyar, E. F. El-Saadany, and M. Sanaye-Pasand, "Fault type classification in microgrids including photovoltaic DGs," *IEEE Trans. Smart Grid*, vol. 7, no. 5, pp. 2218-2229, Sep. 2016.

[39] A. M. Tsimtsios and V. C. Nikolaidis, "Setting zero-sequence compensation factor in distance relays protecting distribution systems," *IEEE Trans. Power Del.*, vol. 33, no. 3, pp. 1236-1246, Jun. 2018.

[40] *IEEE Recommended Practice for Monitoring Electric Power Quality*, IEEE Standard 1159, 2009.

[41] V. C. Nikolaidis, A. D. Patsidis, and A. M. Tsimtsios, "High impedance fault modelling and application of detection techniques with EMTP-RV," *J. Eng.*, vol. 2018, no. 15, pp. 1120-1124, Oct. 2018.

[42] *IEEE Guide for Breaker Failure Protection of Power Circuit Breakers*, IEEE Standard C37.119, 2016.

[43] C. Gellings, "Estimating the costs and benefits of the smart grid: A preliminary estimate of the investment requirements and the resultant benefits of a fully functioning smart grid," Electric Power Research Institute, Palo Alto, CA, USA, Rep. 1022519, 2011.

[44] C. A. P. Meneses and J. R. S. Mantovani, "Improving the grid operation and reliability cost of distribution systems with dispersed generation," *IEEE Trans. Power Syst.*, vol. 28, no. 3, pp. 2485-2496, Aug. 2013.

[45] M. E. Samper and A. Vargas, "Investment decisions in distribution networks under uncertainty with distributed generation—Part I: Model formulation," *IEEE Trans. Power Syst.*, vol. 28, no. 3, pp. 2331-2340, Aug. 2013.

[46] University of Washington, "Power systems test case archive," Seattle, WA, USA, Aug. 1993. [Online]. Available: http://labs.ece.uw.edu/pstca/pf14/pg_tca14bus.htm

## VIII. BIOGRAPHIES

**Aristotelis M. Tsimtsios** (S'17) received the Diploma of Electrical and Computer Engineering and the M.Sc. in Energy Systems and Renewable Energy Sources from the Department of Electrical and Computer Engineering, Democritus University of Thrace, Xanthi, Greece, in 2013 and 2015, respectively. He is now pursuing a Ph.D. at the same Department. His research interests include power system protection/reliability and distributed generation.

**Vassilis C. Nikolaidis** (M' 2011, SM' 2018) received the five-year Diploma of Electrical and Computer Engineering from the Department of Electrical and Computer Engineering, Democritus University of Thrace, Xanthi, Greece, in 2001, the M.Eng. degree in Energy Engineering and Management from National Technical University of Athens (NTUA), Athens, Greece, in 2002, and the Doctor of Engineering from NTUA, in 2007. Since 2008 he has been working as a power systems consulting engineer. Currently he is an Assistant Professor in the Department of Electrical and Computer Engineering, Democritus University of Thrace, Greece. His research interests mainly deal with power system protection, control, stability, and transients.